\documentclass[submission,narroweqnarray,inline]{IEEEtran}
\usepackage {times}
\usepackage{ieeefig}
\usepackage{epsf}
\usepackage[dvips]{epsfig}
\usepackage{pslatex}
\usepackage{multirow}
\usepackage{verbatim}
\usepackage{algorithmic} 
\usepackage[section]{algorithm}
 
\begin{document}
\title{Maximum Error Modeling for Fault-Tolerant Computation using Maximum {\it a posteriori} (MAP) Hypothesis}

\author{
Karthikeyan Lingasubramanian$^{+}$, Syed M. Alam$^{*}$, Sanjukta Bhanja$^{+}$\\
$^{+}${\it Nano Computing Research Group (NCRG), Department of Electrical Engineering, University of South Florida}\\
$^{*}${\it EverSpin Technologies}
}
\maketitle

\begin{center}
{\large \bf Abstract} 
\end{center}

The application of current generation computing machines in safety-centric applications like implantable biomedical chips and automobile safety has immensely increased the need for reviewing the worst-case error behavior of computing devices for fault-tolerant computation. In this work, we propose an exact probabilistic error model that can compute the {\it maximum} error over all possible input space in a circuit-specific manner and can handle various types of structural dependencies in the circuit. We also provide the worst-case input vector, which has the highest probability to generate an erroneous output, for any given logic circuit. We also present a study of circuit-specific error bounds for fault-tolerant computation in heterogeneous circuits using the maximum error computed for each circuit. We model the error
estimation problem as a maximum {\it a posteriori} (MAP) estimate, over
the joint error probability function of the entire circuit, calculated efficiently through an intelligent search of the entire input space using probabilistic traversal of a binary join tree using {\it Shenoy-Shafer} algorithm. We demonstrate this model using MCNC and ISCAS benchmark circuits and validate it using an equivalent HSpice model. Both results yield the same worst-case input vectors and the highest \% difference of our error model over HSpice is just $1.23\%$. We observe that the maximum error probabilities are
significantly larger than the average error probabilities, and provides a much tighter error bounds for fault-tolerant computation. We also find
that the error estimates depend on the specific circuit structure and the maximum error probabilities are sensitive to the individual gate failure probabilities.

\section{Introduction}
\label{introduction}

{\bf Why maximum error?} Industries like automotive and health care, which employs safety-centric electronic devices, have traditionally addressed high
reliability requirements by employing redundancy, error corrections,
and choice of proper assembly and packaging technology. In addition,
rigorous product testing at extended stress conditions filters out
even an entire lot in the presence of a small number of failures~\cite{mason}.
Another rapidly growing class of electronic chips where reliability is
very critical in implantable biomedical chips~\cite{gerrish,stotts}. More
interestingly, some of the safety approaches, such as redundancy and
complex packaging, are not readily applicable to implantable
biomedical applications because of {\it low voltage, low power operation
and small form factor requirements.} Also in future technologies like NW-FET, CNT-FET~\cite{cnt}, RTD~\cite{mazumder}, hybrid nano devices~\cite{jha}, single electron tunneling devices~\cite{chen_mao}, field coupled computing devices like QCA's~\cite{qca} (molecular and
magnetic) and spin-coupled computing devices, computing components are
likely to have higher error rates (both in terms of defect and
transient faults) since they {\it operate near the thermal limit and
information processing occurs at extremely small volume.} Nano-CMOS,
beyond 22nm, is not an exception in this regard as the frequency
scales up and voltage and geometry scales down. Also we have to note that, {\it while two design implementation choices can
have different average probabilities of failures, the lower average
choice may in fact have higher maximum probability of failure leading
to lower yield in manufacturing and more rejects during chip burn-in
and extended screening.} 

\subsection{Proposed Work}
\label{proposed}

In this work, we present a probabilistic model to study the {\it maximum}
output error over all possible input space for a given logic circuit. We present a method to find out the worst-case input vector, i.e., the input vector that has the highest probability to give an error at the output. In the first step of our model, we convert the circuit into a corresponding {\it edge-minimal} probabilistic network that represents the basic logic function of the circuit by handling the interdependencies between the signals using random variables of interest in a composite joint probability distribution function $P(y_{1},y_{2},\cdots,y_{N})$. Each node in this network corresponds to a random variable representing a signal in the digital circuit, and each edge corresponds to the logic governing the connected signals. The individual probability distribution for each node is given using conditional probability tables. 

From this probabilistic network we obtain our probabilistic error model that consists of three blocks,
(i) ideal error free logic, (ii) error prone logic where every gate has a gate error probability $\varepsilon$ i.e., each gate can go wrong individually by a probabilistic factor $\varepsilon$ and (iii) a detection unit that uses comparators to compare the error free and erroneous outputs. The error prone logic represents the real time circuit under test, whereas the ideal logic and the detection unit are fictitious elements used to study the circuit. 
Both the ideal logic and error prone logic would be fed by the primary inputs ${\bf I}$. We denote all the internal nodes, both in the error free and erroneous portions, by ${\bf  X}$ and the comparator outputs as ${\bf O}$. The comparators are based
on XOR logic and hence a state "1" would signify error at the output. An evidence set ${\bf o}$ is created by evidencing one or more of the variables in the comparator set ${\bf O}$ to state "1" ($P(O_{i}=1)=1$). Then performing MAP hypothesis on the probabilistic error model provides the worst-case input vector ${\bf i}_{MAP}$ which gives $\max_{\forall {\bf i}}P({\bf i},{\bf o})$. The maximum output error probability can be obtained from $P(O_{i}=1)$ after instantiating the input nodes of probabilistic error model with ${\bf i}_{MAP}$ and inferencing. The process is repeated for increasing $\varepsilon$ values and finally the $\varepsilon$ value that makes at least one of the output signals completely random ($P(O_{i}=0)=0.5, P(O_{i}=1)=0.5$) is taken as the error bound for the given circuit.

It is obvious that we can arrive at MAP estimate by enumerating all possible input instantiations and compute the maximum $P({\bf i},{\bf o})$ by any probabilistic computing tool. The attractive feature of this MAP algorithm lies  on eliminating a significant part of the input search-subtree based on an easily available upper-bound of $P({\bf i},{\bf o})$ by using probabilistic traversal of a binary Join tree with {\it Shenoy-Shafer} algorithm~\cite{shenoy-shafer,binary-jt}. The actual computation is divided into two theoretical components. First, we convert the circuit structure into a binary Join tree and employ Shenoy-Shafer algorithm, which is a two-pass probabilistic message-passing algorithm, to obtain multitude of upper bounds of $P({\bf i},{\bf o})$ with partial input instantiations. Next, we construct a Binary tree of the input vector space where each path from the root node to the leaf node represents an input vector. At every node, we traverse the search tree if the upper bound, obtained by Shenoy-Shafer inference on the binary join tree, is greater than the maximum probability already achieved; otherwise we prune the entire sub-tree.
Experimental results on a few standard benchmark show that the worst-case errors significantly deviate from the average ones and also provides tighter bounds for the ones that use homogeneous gate-type (c17 with NAND-only). Salient features and deliverables are itemized below:

\begin{itemize}

\item We have proposed a method to calculate {\it maximum} output error using a probabilistic model. Through experimental results, we show the importance of modeling maximum output error. (Fig.~\ref{bounds})

\item Given a circuit with a fixed gate error probability $\varepsilon$, our model can provide the maximum output error probability and the {\it worst-case} input vector, which can be very useful testing parameters.

\item We present the circuit-specific error bounds for fault-tolerant computation and we show that maximum output errors provide a tighter bound.

\item We have used an efficient design framework that employs inference in binary join trees using Shenoy-Shafer algorithm to perform MAP hypothesis accurately.   

\item We give a probabilistic error model, where efficient error incorporation is possible, for useful reliability studies. Using our model the error injection and probability of error for each gate can be modified easily. Moreover, we can accommodate both fixed and variable gate errors in a single circuit without affecting computational complexity.

\end{itemize}

The rest of the paper is structured as follows, Section.~\ref{prior} gives a summary of some of the previous works on error bounds for fault-tolerant computation along with some of the reliability models established from these works, Section.~\ref{model} explains the structure of our probabilistic error model, Section.~\ref{map_theory} explains the MAP hypothesis and its complexity, Section.~\ref{results} provides the experimental results,  followed by conclusion in Section.~\ref{conclusion}. 

\begin{figure*}
\vspace*{-0.3in}
\begin{center}
\epsfxsize 420pt
{\epsffile{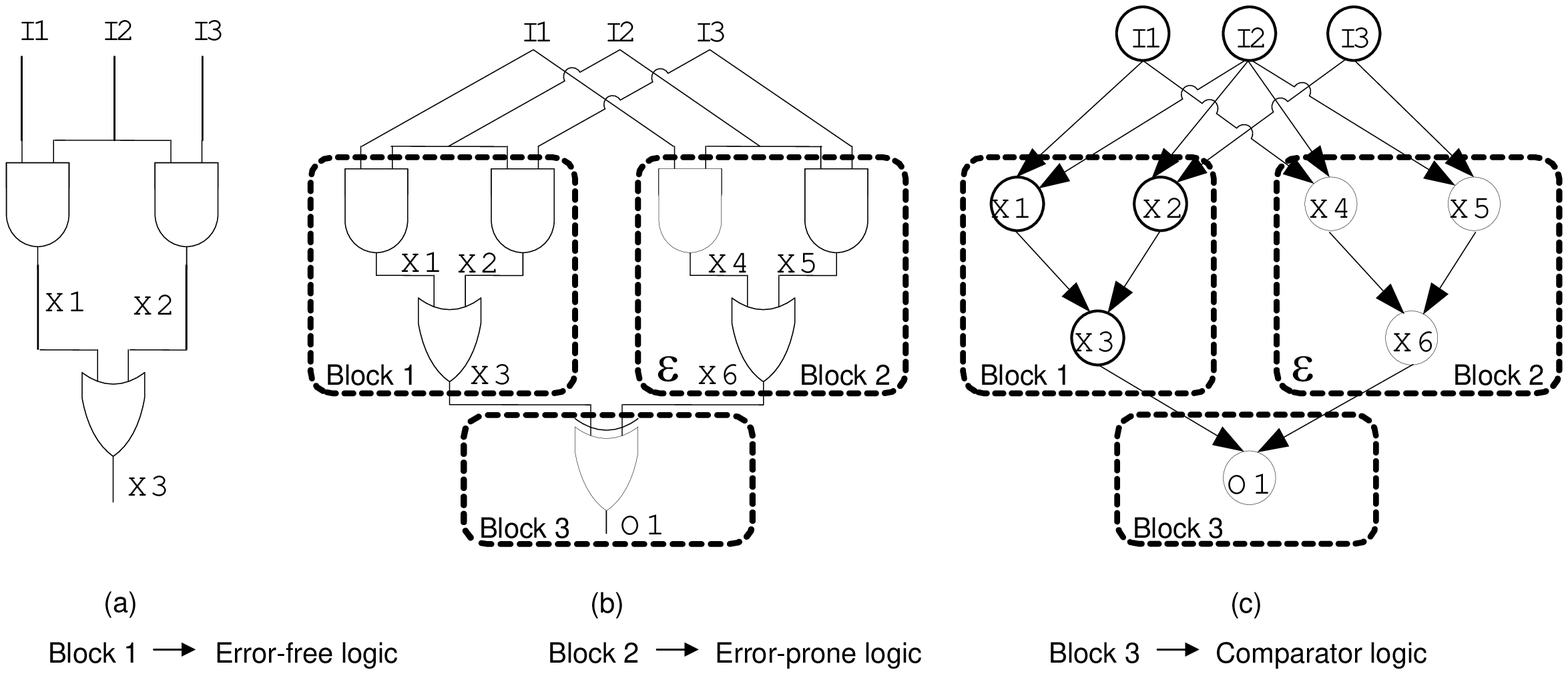}}
\end{center}
\vspace*{-0.2in}
\caption{(a) Digital logic circuit (b) Error model (c) Probabilistic error model}
\label{model_fig}
\end{figure*}
 
\section{Prior Work}
\label{prior}

\subsection{State-of-the-art} 

The study of reliable computation using unreliable components was initiated by von Neumann~\cite{vn} who showed that erroneous components with some small error probability can provide reliable outputs and this is possible only when the error probability of each component is less than $1/6$. This work was later enhanced by Pippenger~\cite{pippenger} who realized von Neumann's model using formulas for Boolean functions. This work showed that for a function controlled by $k$-arguments the error probability of each component should be less than $(k-1)/2k$ to achieve reliable computation. This work was later extended by using networks instead of formulas to realize the reliability model~\cite{feder}. In~\cite{hajek}, Hajek and Weller used the concept of formulas to show that for 3-input gates the error probability should be less than $1/6$. Later this work was extended for $k$-input gates~\cite{evans-schulman} where $k$ was chosen to be odd. For a specific even case, Evans and Pippenger~\cite{evans} showed that the maximum tolerable noise level for 2-input NAND gate should be less than $(3 - \sqrt{7})/4 = 0.08856\cdots$. Later this result was reiterated by Gao {\it et al} for 2-input NAND gate, along with other results for $k$-input NAND gate and majority gate, using bifurcation analysis~\cite{jose02} that involves repeated iterations on a function relating to the specific computational component. While there exists studies of circuit-specific bounds for circuit characteristics like switching activity~\cite{marculescu}, the study of circuit-specific error bounds would be highly informative and useful for designing high-end computing machines.

The study of fault-tolerant computation has expanded its barriers and is being generously employed in fields like nano-computing architectures. Reliability models like Triple Modular Redundancy (TMR) and N-Modular Redundancy (NMR)~\cite{depledge} were designed using the von Neumann model. Expansion of these techniques led to models like Cascaded Triple Modular Redundancy (CTMR)~\cite{spagocci} used for nanochip devices. In~\cite{han}, the reliability of reconfigurable architectures was obtained using NAND multiplexing technique and in~\cite{beiu}, majority multiplexing was used to achieve fault-tolerant designs for nanoarchitectures. A recent comparative study of these methods~\cite{Nikolic02}, indicates that a 1000-fold redundancy would be required for a device error (or failure) rate of 0.01\footnote{Note that this does {\it not} mean 1 out of 100 devices will fail, it indicates the devices will generate erroneous output 1 out of 100 times.}. Many researchers are currently focusing on computing the average  error~\cite{mohanram,beiu-ibrahim} from a circuit and also on the expected error to conduct reliability-redundancy trade-off studies. An approximate method based on Probabilistic Gate Model (PGM) is discussed by Han {\it et al.} in~\cite{jose04}. Here the PGMs are formed using equations governing the functionality between an input and an output. Probabilistic analysis of digital logic circuits using decision diagrams is proposed in ~\cite{abdollahi}. In~\cite{thara}, the average output error in digital circuits is calculated using a probabilistic reliability model that employs Bayesian Networks. 

In testing, the identification of possible input patterns to perform efficient circuit testing is achieved through Automatic Test Pattern Generation (ATPG) algorithms. Some of the commonly used ATPG algorithms like D-algorithm~\cite{d-alg}, PODEM (path-oriented decision making) algorithm~\cite{podem} and FAN (fanout-oriented test generation) algorithm~\cite{fan} are deterministic in nature. There are some partially probabilistic ATPG algorithms~\cite{prob-atpg,savir-atpg,seth-atpg} which are basically used to reduce the input pattern search space. In order to handle transient errors occurring in intermediate gates of a circuit, we need a completely probabilistic model~\cite{neural-atpg}.

\subsection{Relation to State-of-the-art} 

Our work concentrates on {\it estimation of maximum error as opposed to average error}, since for higher design levels it is important to account for maximum error behavior, especially if this behavior is far worse than the average case behavior.

Also our work proposes {\it a completely probabilistic model as opposed to a deterministic model}, where every gate of the circuit is modeled probabilistically and the worst case input pattern is obtained. 

The bounds presented in all the above mentioned works do not consider (i) combination of different
logic units like NAND and majority in deriving the bounds and (ii) do
not consider circuit structure and dependencies and error masking that
could occur in a realistic logic network, making the bounds pessimistic. 
Our model {\it encapsulates the entire circuit structure} along with the signal inter dependencies and so is capable of estimating the error bound of the entire circuit as opposed to a single logic unit.

\section{Probabilistic error model}
\label{model}

The underlying model compares error-free and
error-prone outputs.  Our model contains three sections, (i)
error-free logic where the gates are assumed to be perfect, (ii)
error-prone logic where each gate goes wrong independently by an error
probability $\varepsilon$ and (iii) XOR-logic based comparators that
compare the error-free and error-prone primary outputs. When error occurs, the error-prone primary output signal will not be at the same state as the ideal error-free primary output signal. So, {\it an output of logic "1" at the XOR comparator gate indicates occurrence of error}. For a given digital logic circuit as in Fig.~\ref{model_fig}(a), the error model and the corresponding probabilistic error model are illustrated in Fig.~\ref{model_fig}(b) and Fig.~\ref{model_fig}(c) respectively. In Fig.~\ref{model_fig}(b) and Fig.~\ref{model_fig}(c), block 1 is the error-free logic, block 2 is the error-prone logic with gate error probability $\varepsilon$ and block 3 is the comparator logic. In the entire model, the error-prone portion given in block 2 is the one that represents the real-time circuit. The ideal error-free portion in block 1 and the comparator portion in block 3 are fictitious and used for studying the given circuit. 

We would like the readers to note that we will be representing a {\bf SET OF VARIABLES} by bold capital letters, {\bf set of instantiations} by bold small letters, any SINGLE VARIABLE by capital letters. Also probability of the event $Y_{i}=y_{i}$ will be denoted simply by $P(y_{i})$ or by $P(Y_{i}=y_{i})$.

The probabilistic network is a conditional factoring of a joint probability
distribution. The nodes in the network are random variables representing each signal in the underlying circuit. To perfectly represent digital signals each random variable will have two states, state $"0"$ and state $"1"$. The edges represent the logic that governs the connecting nodes using conditional probability tables (CPTs). For example, in Fig.~\ref{model_fig}(c), the nodes $X1$ and $X4$ are random variables representing the error-free signal $X1$ and the error-prone signal $X4$ respectively of the digital circuit given in Fig.~\ref{model_fig}(a). The edges connecting these nodes to their parents $I1$ and $I2$ represent the error-free AND logic and error-prone AND logic as given by the CPTs in Table.~\ref{cpt}. 

\begin{table} 
\caption{Conditional Probability Tables (CPTs) for error-free and error-prone AND logic} 
\label{cpt} 
\begin{center} 
\begin{tabular}{||c|c|c||} \hline  
\multicolumn{3}{|c|}{Error-free AND} \\ \hline
$P(X1=1|I1,I2)$ & $P(I2=0)=1$ & $P(I2=1)=1$  \\ \hline 
$P(I1=0)=1$ & 0 & 0 \\ \hline 
$P(I1=1)=1$ & 0 & 1 \\ \hline 
\multicolumn{3}{|c|}{Error-prone AND} \\ \hline
$P(X4=1|I1,I2)$ & $P(I2=0)=1$ & $P(I2=1)=1$  \\ \hline 
$P(I1=0)=1$ & $\varepsilon$ & $\varepsilon$ \\ \hline 
$P(I1=1)=1$ & $\varepsilon$ & 1-$\varepsilon$ \\ \hline 
\end{tabular}  
\end{center}  
\end{table} 

Let us define the random variables in our probabilistic error model as ${\bf Y} = {\bf I} \cup {\bf X} \cup {\bf O}$, composed of the three disjoint subsets ${\bf I}$, ${\bf X}$ and ${\bf O}$ where

\begin{enumerate}

\item $I_{1},\cdots,I_{k} \in {\bf I}$ are the
set of $k$ primary inputs.
\item $X_{1},\cdots,X_{m} \in {\bf X}$ are the $m$ internal logic signals for both the erroneous (every gate has a
failure probability $\varepsilon$) and error-free ideal logic elements.
\item $O_{1},\cdots,O_{n} \in {\bf O}$ are the $n$ comparator outputs,
each one signifying the error in one of the primary outputs of the
logic block. 
\item $N=k+m+n$ is the total number of network random
variables.
\end{enumerate}

Any probability function
$P(y_{1},y_{2},\cdots, y_{N})$, where $y_{1},y_{2},\cdots, y_{N}$ are random variables, can be written as,
\begin{eqnarray}
  P(y_{1}, \cdots, y_{N}) &=& P(y_{N}|y_{N-1}, y_{N-2},\cdots,y_{1}) \nonumber \\ 
  & & P( y_{N-1}|y_{N-2}, y_{N-3},\cdots,y_{1}) \nonumber \\
  & &\cdots P(y_{1})
\label{joint-prob_1}
\end{eqnarray}
This expression holds for any ordering of the random variables. In
most applications, a variable is usually not dependent on all other
variables. There are lots of conditional independencies embedded among
the random variables, which can be used to reorder the random
variables and to simplify the joint probability as,
\begin{equation}
P(y_{1}, \cdots, y_{N}) = \prod_{v} P(y_{v}|Pa(Y_{v})) 
\label{joint-prob}
\end{equation}
where $Pa(Y_{v})$ indicates the parents of the variable $Y_{v}$, representing
its direct causes. This factoring of the joint probability function
can be denoted as a graph with links directed from the random variable representing the inputs of a gate to the random variable representing the output. To understand it better let us look at the error model given in Fig.~\ref{model_fig}(c). The joint probability distribution representing the network can be written as,
\begin{eqnarray}
  P(i1,i2,i3,x1,\cdots,x6,o1) &=& P(o1|x6,\cdots,x1,i3,i2,i1) \nonumber \\ 
  & & P(x6|x5,\cdots,x1,i3,i2,i1) \nonumber \\
  & &\cdots P(i3)P(i2)P(i1)
\label{joint-prob_exp}
\end{eqnarray}
Here the random variable $O1$ is independent of the random variables $X1,X2,X4,X5,I1,I2,I3$ given its parents $X3,X6$. This notion explains the conditional independence between the random variables in the network and it is mathematically denoted by $I(O1,\{X3,X6\},\{X1,X2,X4,X5,I1,I2,I3\})$. So for $O1$, the probability distribution can be rephrased as,
\begin{equation}
  P(o1|x6,\cdots,x1,i3,i2,i1) = P(o1|x6,x3)
\label{joint-prob_exp1}
\end{equation}
By implementing all the underlying conditional independencies the basic joint probability distribution can be rephrased as,
\begin{eqnarray}
  P(i1,i2,i3,x1,\cdots,x6,o1) &=& P(o1|x6,x3)P(x6|x5,x4) \nonumber \\
  & & P(x5|i3,i2)P(x4|i2,i1) \nonumber \\
  & & P(x3|x2,x1)P(x2|i3,i2) \\
  & & P(x1|i2,i1)P(i3)P(i2)P(i1) \nonumber 
\label{joint-prob_exp2}
\end{eqnarray}

The implementation of this probability distribution can be clearly seen in Fig.~\ref{model_fig}(c). Each node is connected only to its parents and not to any other nodes. The conditional probability potentials for all the nodes are provided by the CPTs. The attractive feature of this graphical representation of the joint probability distribution is that
not only does it make conditional dependency relationships among the nodes explicit but it also serve as a computational mechanism for efficient probabilistic updating. 

\section{Maximum {\it a Posteriori} (MAP) Estimate}
\label{map_theory}
\begin{figure*}
\vspace*{-1.0in}
\begin{center}
\epsfxsize 450pt
\epsffile{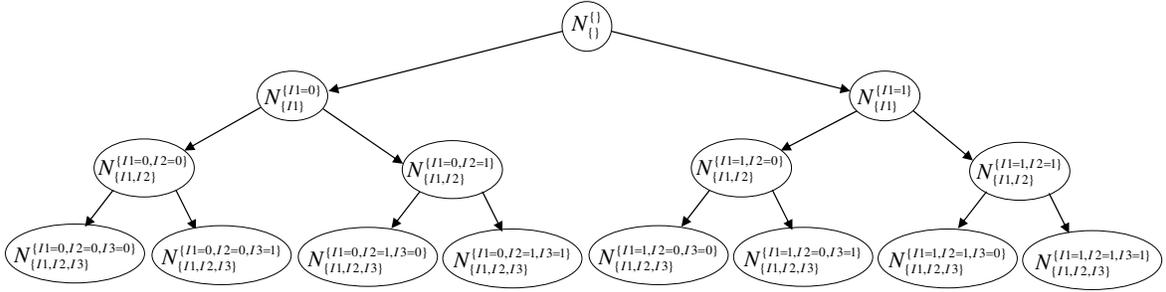}
\end{center}
\vspace*{-1.9in}
\caption{Search tree where depth first branch and bound search performed.}
\label{map_trees}
\end{figure*}

As we mentioned earlier, in our probabilistic error model, the network variables ,say ${\bf Y}$, can be divided into three subsets ${\bf I}$, ${\bf X}$ and ${\bf O}$ where $I_{1},\cdots,I_{k} \in {\bf I}$ represents primary input signals; $X_{1},\cdots,X_{m} \in {\bf X}$ represents internal signals including the primary output signals; $O_{1},\cdots,O_{n} \in {\bf O}$ represents the comparator output signals. {\it Any primary output node can be forced to be erroneous by fixing the corresponding comparator output to logic "1"}, that is providing an {\it evidence} ${\bf o} = \{P(O_{i}=1)=1\}$ to a comparator output $O_{i}$. Given some evidence ${\bf o}$, the objective of the Maximum {\it a posteriori} estimate is to find a complete instantiation ${\bf i}_{MAP}$ of the variables in ${\bf I}$ that gives the following joint probability, 
\begin{equation}
MAP({\bf i}_{MAP},{\bf o}) = \max_{\forall{\bf i}} P({\bf i},{\bf o})
\end{equation}
The probability $MAP({\bf i}_{MAP},{\bf o})$ is termed as the {\it MAP probability} and the variables in ${\bf I}$ are termed as {\it MAP variables} and the instantiation ${\bf i}_{MAP}$ which gives the maximum $P({\bf i},{\bf o})$ is termed as the {\it MAP instantiation}. 

For example, consider Fig~\ref{model_fig}. In the probabilistic model shown in
Fig~\ref{model_fig}(c), we have $\{I1,I2,I3\} \in {\bf I}$;
$\{X1,X2,X3,X4,X5,X6\} \in {\bf X}$; $\{O1\} \in {\bf O}$. $X3$ is the ideal error-free primary output node and $X6$ is the corresponding error-prone primary output node. Giving an evidence ${\bf o} = \{P(O1=1)=1\}$ to $O1$ indicates that $X6$ has produced an erroneous output. The MAP hypothesis uses this information and finds the input instantiation, ${\bf i}_{MAP}$, that would give the maximum $P({\bf i},{\bf o})$. This indicates that ${\bf i}_{MAP}$ is the most probable input instantiation that would give an error in the error-prone primary output signal $X6$. In this case, ${\bf i}_{MAP}=\{I1=0,I2=0,I3=0\}$. This means that the input instantiation $\{I1=0,I2=0,I3=0\}$ will most probably provide a wrong output, $X6=1$ (since the correct output is $X6=0$).

We arrive at the exact Maximum {\it a posteriori} (MAP) estimate using the algorithms by Park and Darwiche~\cite{park_exact}~\cite{hill_climb}. It is obvious that we could arrive at MAP estimate by enumerating all possible input instantiations and compute the maximum output error. 
To make it more efficient, our MAP estimates rely on eliminating some part of the input search-subtree based on an easily available upper-bound of MAP probability by using a probabilistic traversal of a binary Join tree using {\it Shenoy-Shafer} algorithm~\cite{shenoy-shafer,binary-jt}. The actual computation is divided into two theoretical components.

\begin{itemize}
\item First, we convert the circuit structure into a binary Join tree and employ Shenoy-Shafer algorithm, which is a two-pass probabilistic message-passing algorithm, to obtain multitude of upper bounds of MAP probability with partial input instantiations (discussed in Section.~\ref{ss-algorithm}). The reader familiar with Shenoy-Shafer algorithm can skip the above section. To our knowledge, Shenoy-Shafer algorithm is not commonly used in VLSI context, so we elaborate most steps of join tree creation, two-pass join tree traversal and computation of upper bounds with partial input instantiations. 

\item Next, we construct a Binary tree of the input vector space where each path from the root node to the leaf node represents an input vector. At every node, we traverse the search tree if the upper bound, obtained by Shenoy-Shafer inference on the binary join tree, is greater than the maximum probability already achieved; otherwise we prune the entire sub-tree.  The depth-first traversal in the binary input instantiation tree is discussed in Section.~\ref{final_map} where we detail the search process, pruning and heuristics used for better pruning.  Note that the pruning is key to the significantly improved efficiency of the MAP estimates.
\end{itemize}

\subsection{Calculation of MAP upper bounds using Shenoy-Shafer algorithm}
\label{ss-algorithm}

To clearly understand the various MAP probabilities that are calculated during MAP hypothesis, let us see the binary search tree formed using the MAP variables. A complete search through the MAP variables can be illustrated as shown in Fig.~\ref{map_trees} which gives the corresponding search tree for the probabilistic error model given in Fig.~\ref{model_fig}(c). In this search tree, the root node $N$ will have an empty instantiation; every intermediate node $N_{{\bf I}_{inter}}^{{\bf i}_{inter}}$ will be associated with a subset ${\bf I}_{inter}$ of MAP variables ${\bf I}$ and the corresponding partial instantiation ${\bf i}_{inter}$; and every leaf node $N_{{\bf I}}^{{\bf i}}$ will be associated with the entire set ${\bf I}$ and the corresponding complete instantiation ${\bf i}$. Also each node will have $v$ children where $v$ is the number of values or states that can be assigned to each variable $I_{i}$. Since we are dealing with digital signals, every node in the search tree will have two children. Since the MAP variables represent the primary input signals of the given digital circuit, one path from the root to the leaf node of this search tree gives one input vector choice. In Fig.~\ref{map_trees}, at node $N_{\{I1,I2\}}^{01}$, ${\bf I}_{inter} = \{I1,I2\}$ and ${\bf i}_{inter} = \{I1=0, I2=1\}$. The basic idea of the search process is to find the MAP probability $MAP({\bf i},{\bf o})$ by finding the upper bounds of the intermediate MAP probabilities $MAP({\bf i}_{inter},{\bf o})$.

MAP hypothesis can be categorized into two portions. The first portion involves finding intermediate {\it upper bounds} of MAP probability, $MAP({\bf i}_{inter},{\bf o})$, and the second portion involves {\it improving} these bounds to arrive at the exact MAP solution, $MAP({\bf i}_{MAP},{\bf o})$. These two portions are intertwined and performed alternatively to effectively improve on the intermediate MAP upper bounds. These upper bounds and final solution are calculated by performing inference on the probabilistic error model using Shenoy-Shafer algorithm~\cite{shenoy-shafer,binary-jt}. 

Shenoy-Shafer algorithm is based on local computation mechanism. The probability distributions of the locally connected variables are propagated to get the joint probability distribution of the entire network from which any individual or joint probability distributions can be calculated. The Shenoy-shafer algorithm involves the following crucial information and calculations. 

\begin{table*} 
\caption{Valuations of the variables derived from corresponding CPTs} 
\label{valuations} 
\begin{center} 
\tiny{
\begin{tabular}{c c c c}
CPT
&
\begin{tabular}{||c|c|c||} \hline  
\multicolumn{3}{|c|}{Error-free AND} \\ \hline
$P(X1=1|I1,I2)$ & $P(I2=0)=1$ & $P(I2=1)=1$  \\ \hline 
$P(I1=0)=1$ & 0 & 0 \\ \hline 
$P(I1=1)=1$ & 0 & 1 \\ \hline 
\end{tabular}
&
\begin{tabular}{||c|c|c||} \hline  
\multicolumn{3}{|c|}{Error-prone AND} \\ \hline
$P(X4=1|I1,I2)$ & $P(I2=0)=1$ & $P(I2=1)=1$  \\ \hline 
$P(I1=0)=1$ & $\varepsilon$ & $\varepsilon$ \\ \hline 
$P(I1=1)=1$ & $\varepsilon$ & 1-$\varepsilon$ \\ \hline 
\end{tabular}  
&
\begin{tabular}{||c|c||} \hline  
\multicolumn{2}{|c|}{Input} \\ \hline
$P(I1=0)$ & 0.5 \\ \hline 
$P(I1=1)$ & 0.5 \\ \hline 
\end{tabular}  
\\
\\
Valuation
&
\begin{tabular}{||c c c|c||} \hline  
\multicolumn{4}{|c|}{Error-free AND} \\ \hline
$X1$ & $I1$ & $I2$ & $\phi_{X1}$  \\ \hline 
0 & 0 & 0 & 1\\ \hline 
0 & 0 & 1 & 1\\ \hline 
0 & 1 & 0 & 1\\ \hline 
0 & 1 & 1 & 0\\ \hline 
1 & 0 & 0 & 0\\ \hline 
1 & 0 & 1 & 0\\ \hline 
1 & 1 & 0 & 0\\ \hline 
1 & 1 & 1 & 1\\ \hline 
\end{tabular}
&
\begin{tabular}{||c c c|c||} \hline  
\multicolumn{4}{|c|}{Error-prone AND} \\ \hline
$X4$ & $I1$ & $I2$ & $\phi_{X4}$  \\ \hline 
0 & 0 & 0 & 1-$\varepsilon$ \\ \hline 
0 & 0 & 1 & 1-$\varepsilon$ \\ \hline 
0 & 1 & 0 & 1-$\varepsilon$ \\ \hline 
0 & 1 & 1 & $\varepsilon$ \\ \hline 
1 & 0 & 0 & $\varepsilon$ \\ \hline 
1 & 0 & 1 & $\varepsilon$ \\ \hline 
1 & 1 & 0 & $\varepsilon$ \\ \hline 
1 & 1 & 1 & 1-$\varepsilon$ \\ \hline 
\end{tabular}
&
\begin{tabular}{||c|c||} \hline  
\multicolumn{2}{|c|}{Input} \\ \hline
$I1$ & $\phi_{I1}$  \\ \hline 
0 & 0.5 \\ \hline 
1 & 0.5 \\ \hline 
\end{tabular}
\end{tabular}
}  
\end{center}  
\end{table*} 

{\it Valuations}: The valuations are functions based on the prior probabilities of the variables in the network.  A valuation for a variable $Y_{i}$ can be given as $\phi_{Y_{i}}=P(Y_{i},Pa(Y_{i}))$ where $Pa(Y_{i})$ are the parents of $Y_{i}$. For variables without parents, the valuations can be given as $\phi_{Y_{i}}=P(Y_{i})$. These valuations can be derived from the CPTs (discussed in Section.~\ref{model}) as shown in Table~\ref{valuations}.

\begin{table} 
\caption{Combination} 
\label{combinations} 
\begin{center} 
\scriptsize{
\begin{tabular}{c c c}
\begin{tabular}{||c c|c||} \hline  
$x$ & $y$ & $f_{xy}$  \\ \hline 
0 & 0 & 1\\ \hline 
0 & 1 & 1\\ \hline 
1 & 0 & 1\\ \hline 
1 & 1 & 0\\ \hline 
\end{tabular}
&
\begin{tabular}{||c c|c||} \hline  
$y$ & $z$ & $f_{yz}$  \\ \hline 
0 & 0 & 1\\ \hline 
0 & 1 & 0\\ \hline 
1 & 0 & 0\\ \hline 
1 & 1 & 0\\ \hline 
\end{tabular}
&
\begin{tabular}{||c c c|c||} \hline  
$x$ & $y$ & $z$ & $f_{xyz}=f_{xy}\otimes f_{yz}$  \\ \hline 
0 & 0 & 0 & 1x1 \\ \hline 
0 & 0 & 1 & 1x0 \\ \hline 
0 & 1 & 0 & 1x0 \\ \hline 
0 & 1 & 1 & 1x0 \\ \hline 
1 & 0 & 0 & 1x1 \\ \hline 
1 & 0 & 1 & 1x0 \\ \hline 
1 & 1 & 0 & 0x0 \\ \hline 
1 & 1 & 1 & 0x0 \\ \hline 
\end{tabular}
\end{tabular}
}  
\end{center}  
\end{table} 

\begin{figure}
\begin{center}
\epsfxsize 250pt
\epsffile{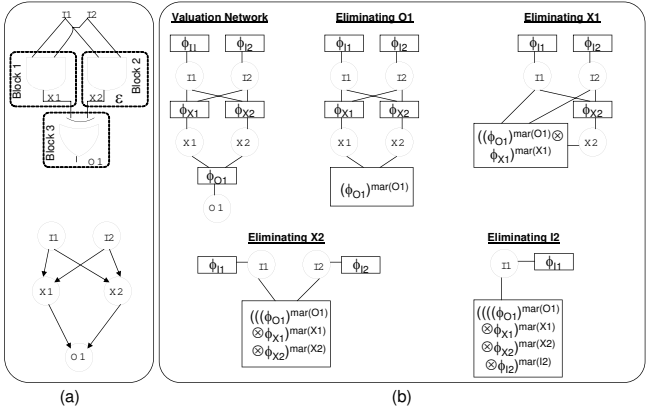}
\end{center}
\vspace*{-0.25in}
\caption{Illustration of the Fusion algorithm.}
\label{fusion-fig}
\end{figure}

{\it Combination}: Combination is a pointwise multiplication mechanism conducted to combine the information provided by the operand functions. A combination of two given functions $f_{a}$ and $f_{b}$ can be written as $f_{a\cup b} = f_{a}\otimes f_{b}$, where $a$ and $b$ are set of variables. Table~\ref{combinations} provides an example.

{\it Marginalization}: Given a function $f_{a\cup b}$, where $a$ and $b$ are set of variables, marginalizing over $b$ provides a function of $a$ and that can be given as $f_{a} = f_{a\cup b}^{mar(b)}$. This process provides the marginals of a single variable or a set of variables. Generally the process can be done by summing or maximizing or minimizing over the {\it marginalizing variables} in $b$. Normally the summation operator is used to calculate the probability distributions. In MAP hypothesis both summation and maximization operators are involved. 

The computational scheme of the Shenoy-Shafer algorithm is based on {\it fusion} algorithm proposed by Shenoy in~\cite{fusion}. Given a probabilistic network, like our probabilistic error model in Fig.~\ref{fusion-fig}(a), the {\it fusion} method can be explained as follows,
\begin{enumerate}

\item The valuations provided are associated with the corresponding variables forming a valuation network as shown in Fig.~\ref{fusion-fig}(b). In our example, the valuations are $\phi_{I1}$ for $\{I1\}$, $\phi_{I2}$ for $\{I2\}$, $\phi_{X1}$ for $\{X1,I1,I2\}$, $\phi_{X2}$ for $\{X2,I1,I2\}$, $\phi_{O1}$ for $\{O1,X1,X2\}$.

\item A variable $Y_{i} \in {\bf Y}$ for which the probability distribution has to be found out is selected. In our example let us say we select $I1$.

\item Choose an arbitrary variable elimination order. For the example network let us choose the order as O1,X1,X2,I2. When a variable $Y_{i}$ is eliminated, the functions associated with that variable $f_{Y_{i}}^{1},\cdots f_{Y_{i}}^{j}$ are combined and the resulting function is marginalized over $Y_{i}$. It can be represented as, $(f_{Y_{i}}^{1}\otimes \cdots \otimes f_{Y_{i}}^{j})^{mar(Y_{i})}$. This function is then associated with the neighbors of $Y_{i}$. This process is repeated until all the variables in the elimination order are removed. Fig.~\ref{fusion-fig} illustrates the fusion process. 

Eliminating $O1$ yields the function $(\phi_{O1})^{mar(O1)}$ associated to neighbors $X1, X2$.\\
Eliminating $X1$ yields the function $((\phi_{O1})^{mar(O1)}\otimes \phi_{X1})^{mar(X1)}$ associated to neighbors $X2, I1, I2$. \\
Eliminating $X2$ yields the function $(((\phi_{O1})^{mar(O1)}\otimes \phi_{X1})^{mar(X1)}\otimes \phi_{X2})^{mar(X2)}$ associated to neighbors $I1, I2$. \\
Eliminating $I2$ yields the function $((((\phi_{O1})^{mar(O1)}\otimes \phi_{X1})^{mar(X1)}\otimes \phi_{X2})^{mar(X2)}\otimes \phi_{I2})^{mar(I2)}$ associated to neighbor $I1$.

According to a theorem presented in~\cite{binary-jt}, combining the functions associated with $I1$ yields the probability distribution of $I1$. $\phi_{I1}\otimes ((((\phi_{O1})^{mar(O1)}\otimes \phi_{X1})^{mar(X1)}\otimes \phi_{X2})^{mar(X2)}\otimes \phi_{I2})^{mar(I2)} = (\phi_{I1}\otimes \phi_{O1}\otimes \phi_{X1}\otimes \phi_{X2}\otimes \phi_{I2})^{mar(O1,X1,X2,I2)}$ = Probability distribution of I1~\cite{binary-jt}. Note that the function $\phi_{I1}\otimes \phi_{O1}\otimes \phi_{X1}\otimes \phi_{X2}\otimes \phi_{I2}$ represents the joint probability of the entire probabilistic error model.

\item The above process is repeated for all the other variables individually.

\end{enumerate}

\begin{figure*}
\begin{center}
\begin{tabular}{c c}
\epsfxsize 200pt
\epsffile{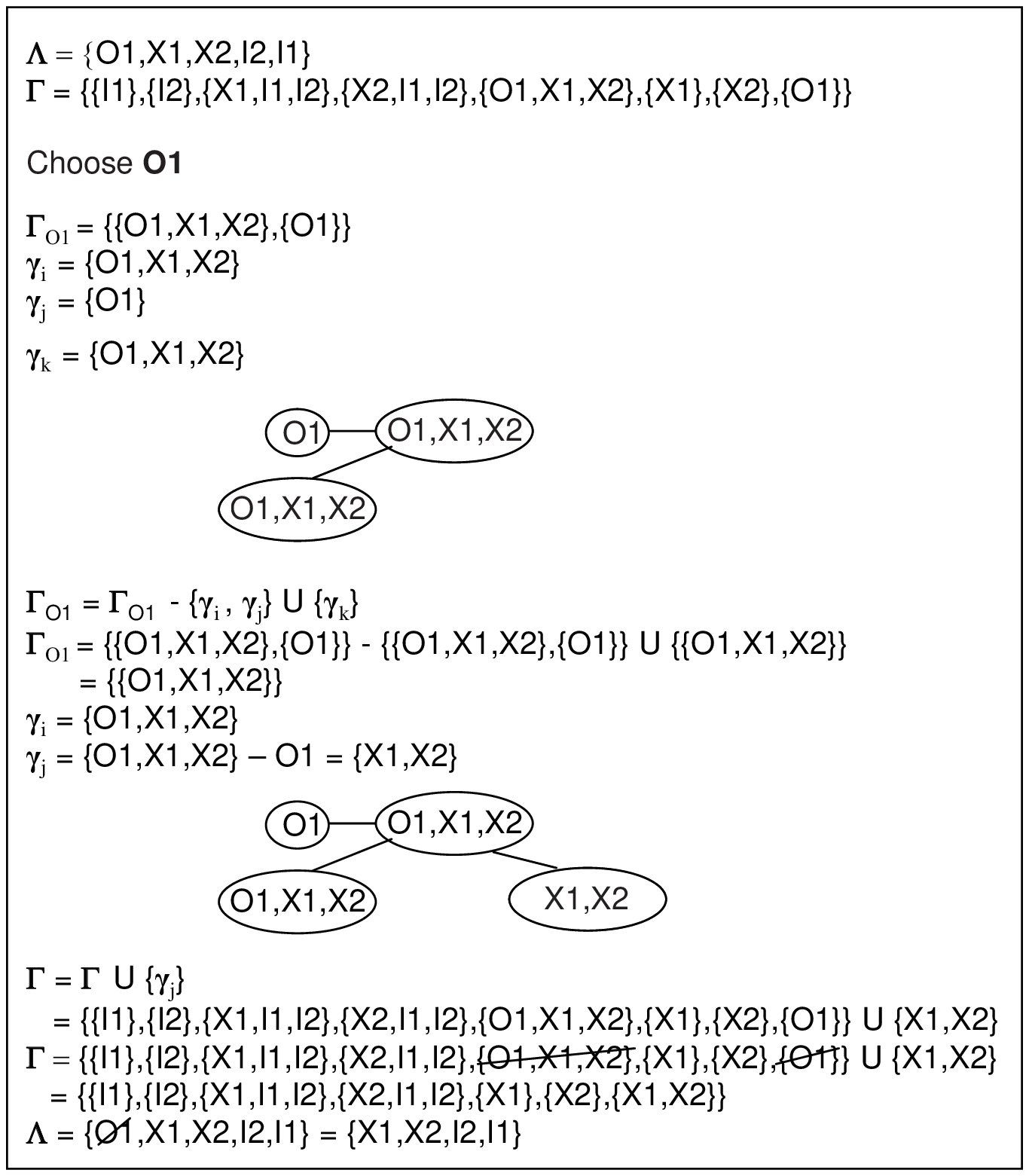}
&
\epsfxsize 300pt
\epsffile{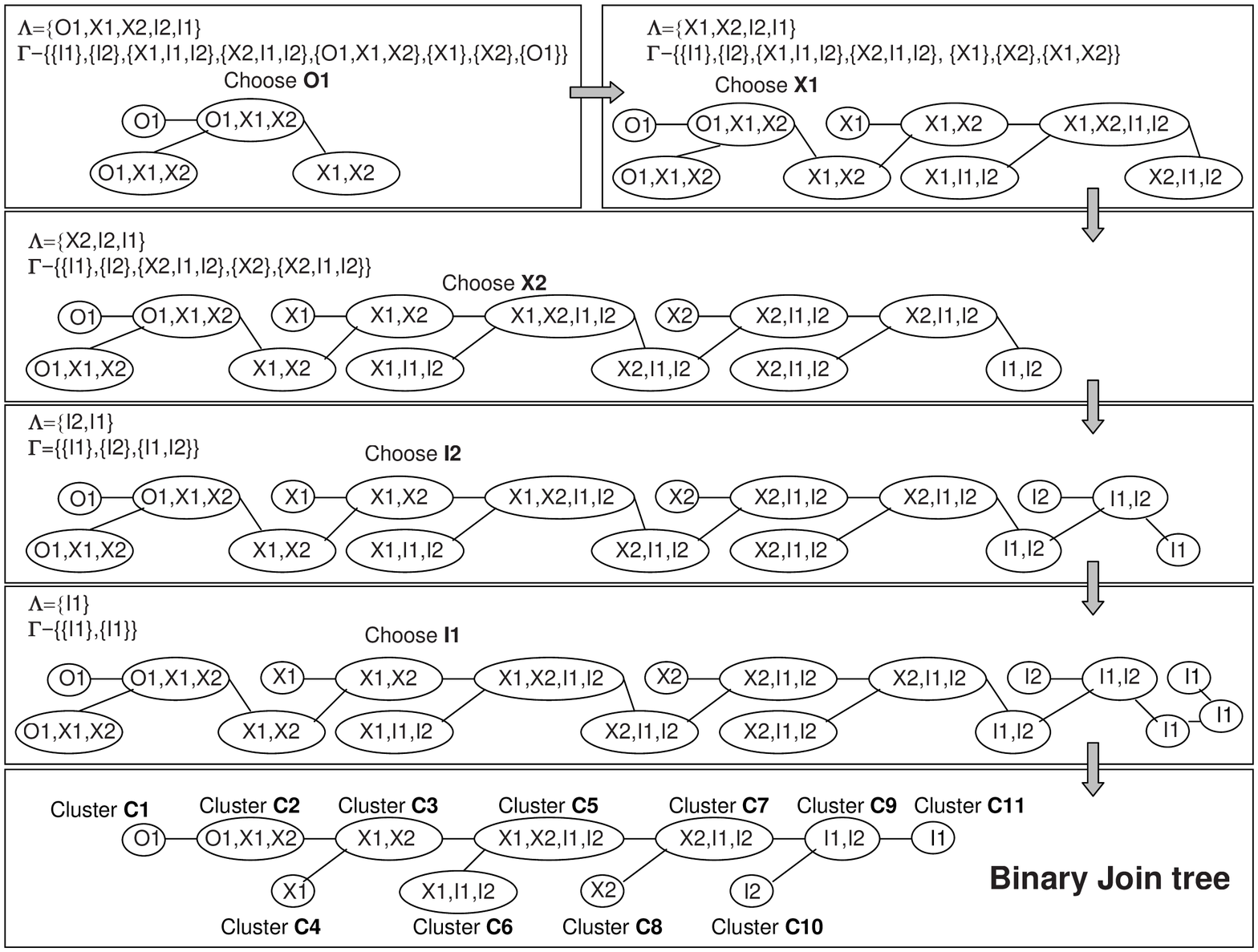}
\\ (a) & (b)
\end{tabular}
\end{center}
\caption{(a) Partial illustration of Binary Join tree construction method for the first chosen variable. (b) Complete illustration of Binary Join tree construction method.}
\label{jointree}
\end{figure*}

To perform efficient computation, an additional undirected network called {\it join tree} is formed from the original probabilistic network. The nodes of the join tree contains {\it clusters} of nodes from the original probabilistic network. The information of locally connected variables, provided through valuations, is propagated in the join tree by {\it message passing} mechanism. 
To increase the computational efficiency of the Shenoy-Shafer algorithm, a special kind of join tree named {\it binary join tree} is used. In a binary join tree, every node is connected to no more than three neighbors. In this framework only two functions are combined at an instance, thereby reducing the computational complexity. We will first explain the method to construct a binary join tree, as proposed by Shenoy in~\cite{binary-jt}, and then we will explain the inference scheme using message passing mechanism. 

{\bf Construction of Binary Join Tree}:
The binary join tree is constructed using the fusion algorithm. The construction of binary join tree can be explained as follows,

\begin{enumerate}

\item To begin with we have,\\
$\Lambda \Longrightarrow$ A set that contains all the variables from the original probabilistic network. In our example, $\Lambda = \{I1,I2,X1,X2,O1\}$.\\
$\Gamma \Longrightarrow$ A set that contains the subsets of variables, that should be present in the binary join tree. i.e., the subsets that denote the valuations and the subsets whose probability distributions are needed to be calculated. In our example, let us say that we need to calculate the individual probability distributions of all the variables. Then we have, $\Gamma =$ \{\{I1\}, \{I2\}, \{X1,I1,I2\}, \{X2,I1,I2\}, \{O1,X1,X2\}, \{X1\}, \{X2\}, \{O1\}\}.\\
$\mathcal{N} \Longrightarrow$ A set that contains the nodes of the binary join tree and it is initially null.\\
$\mathcal{E} \Longrightarrow$ A set that contains the edges of the binary join tree and it is initially null.\\
We also need an order in which we can choose the variables to form the binary join tree. In our example, since the goal is to find out the probability distribution of I1, this order should reflect the variable elimination order (O1,X1,X2,I2,I1) used in fusion algorithm .

\item 

\begin{algorithmic}[1]
\WHILE{$|\Gamma|>1$} 
	\STATE Choose a variable $Y \in \Lambda$ 
	\STATE $\Gamma_{Y} = \{\gamma_{i}\in \Gamma|Y\in \gamma_{i}\}$ 
	\WHILE{$|\Gamma_{Y}|>1$}
		\STATE Choose $\gamma_{i}\in \Gamma_{Y}$ and $\gamma_{j}\in \Gamma_{Y}$ such that $||\gamma_{i}\cup \gamma_{j}||\leq ||\gamma_{m}\cup \gamma_{n}||$ for all $\gamma_{m},\gamma_{n}\in \Gamma_{Y}$
		\STATE $\gamma_{k} = \gamma_{i}\cup \gamma_{j}$
		\STATE $\mathcal{N} = \mathcal{N}\cup \{\gamma_{i}\}\cup \{\gamma_{j}\}\cup \{\gamma_{k}\}$
		\STATE $\mathcal{E} = \mathcal{E}\cup \{\{\gamma_{i},\gamma_{k}\},\{\gamma_{j},\gamma_{k}\}\}$
		\STATE $\Gamma_{Y} = \Gamma_{Y} - \{\gamma_{i},\gamma_{j}\}$
		\STATE $\Gamma_{Y} = \Gamma_{Y}\cup \{\gamma_{k}\}$
	\ENDWHILE
	\IF{$|\Lambda|>1$}
		\STATE Take $\gamma_{i}$ where $\gamma_{i}=\Gamma_{Y}$
		\STATE $\gamma_{j} = \gamma_{i} - \{Y\}$
		\STATE $\mathcal{N} = \mathcal{N}\cup \{\gamma_{i}\}\cup \{\gamma_{j}\}$
		\STATE $\mathcal{E} = \mathcal{E}\cup \{\{\gamma_{i},\gamma_{j}\}\}$
		\STATE $\Gamma = \Gamma \cup \{\gamma_{j}\}$
	\ENDIF	 
	\STATE $\Gamma = \Gamma - \{\gamma_{i}\in \Gamma|Y\in \gamma_{i}\}$ 
	\STATE $\Lambda = \Lambda - \{Y\}$	
\ENDWHILE
\end{algorithmic} 

\item The final structure will have some duplicate clusters. Two neighboring duplicate clusters can be merged into one, if the merged node does not end up having more than three neighbors. After merging the duplicate nodes we get the binary join tree.

\end{enumerate}

Fig.~\ref{jointree} illustrates the binary join tree construction method for the probabilistic error model in Fig.~\ref{fusion-fig}(a). Fig.~\ref{jointree}(a) explains a portion of the construction method for the first chosen variable, here it is $O1$. Fig.~\ref{jointree}(b) illustrates the entire method. Note that, even though the binary join tree is constructed with a specific variable elimination order for finding out the probability distribution of I1, it can be used to find out the probability distributions of other variables too.

\begin{figure}
\begin{center}
\epsfxsize 250pt
\epsffile{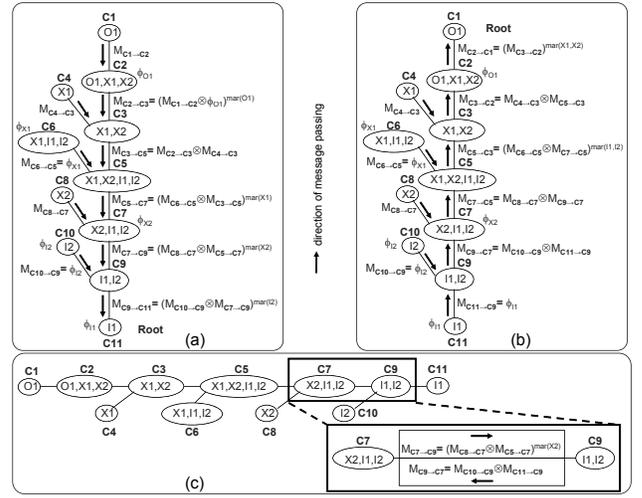}
\end{center}
\caption{(a) Message passing with cluster {\bf C11} as root. (b) Message passing with cluster {\bf C1} as root. (c) Message storage mechanism.}
\vspace*{-0.1in}
\label{messagepassing}
\end{figure}

{\bf Inference in binary join tree}:
Inference in a binary join tree is performed using message passing mechanism. Initially all the valuations are associated to the appropriate clusters. In our example, at Fig.~\ref{messagepassing}, the valuations are associated to these following clusters,\\
- $\phi_{I1}$ associated to cluster {\bf C11}\\
- $\phi_{I2}$ associated to cluster {\bf C10}\\
- $\phi_{X1}$ associated to cluster {\bf C6}\\
- $\phi_{X2}$ associated to cluster {\bf C7}\\
- $\phi_{O1}$ associated to cluster {\bf C2}\\
A message passed from cluster $b$, containing a variable set {\bf B}, to cluster $c$, containing a variable set {\bf C} can be given as,
\begin{equation}
M_{b\rightarrow c} = (\phi_{b} \prod_{a\neq c} M_{a\rightarrow b})^{mar({\bf B}\setminus {\bf C})}
\label{message}
\end{equation}
where $\phi_{b}$ is the valuation associated with cluster $b$. If cluster $b$ is not associated with any valuation, then this function is omitted from the equation. The message from cluster $b$ can be sent to cluster $c$ only after cluster $b$ receives messages from all its neighbors other than $c$. The resulting function is marginalized over the variables in cluster $b$ that are not in cluster $c$. To calculate the probability distribution of a variable $Y_{i}$, the cluster having that variable alone is taken as root and the messages are passed towards this root. Probability of $Y_{i}$, $P(Y_{i})$, is calculated at the root. In our example, at Fig.~\ref{messagepassing}(a), to find the probability distribution of I1, the cluster {\bf C11} is chosen as the root. The messages from all the leaf clusters are sent towards {\bf C11} and finally the probability distribution of I1 can be calculated as, $P(I1) = M_{{\bf C9}\rightarrow{\bf C11}}\otimes\phi_{I1}$. Also note that the {\it order of the marginalizing variables} is O1,X1,X2,I2 which exactly reflects the elimination order used to construct the binary join tree. As we mentioned before, this binary join tree can be used to calculate probability distributions of other variables also. In our example, at Fig.~\ref{messagepassing}(b), to find out the probability distribution of O1, cluster {\bf C1} is chosen as root and the messages from the leaf clusters are passed towards {\bf C1} and finally the probability distribution of O1 can be calculated as, $P(O1) = M_{{\bf C2}\rightarrow{\bf C1}}$. Note that the {\it order of the marginalizing variables} changes to I1,I2,X1,X2. We can also calculate joint probability distributions of the set of variables that forms a cluster in the binary join tree. In our example, the joint probability $P(I1,I2)$ can be calculated by assigning cluster {\bf C9} as root. In this fashion, the probability distributions of any individual variable or a set of variables can be calculated by choosing appropriate root cluster and sending the messages towards this root. During these operations some of the calculations are not modified and so performing them again will prove inefficient. Using the binary join tree structure these calculations can be stored thereby eliminating the redundant recalculation. In the binary join tree, between any two clusters $b$ and $c$, both the messages $M_{b\rightarrow c}$ and $M_{c\rightarrow b}$ are stored. Fig.~\ref{messagepassing}(c) illustrates this phenomenon using our example.

If an evidence set ${\bf e}$ is provided, then the additional valuations $\{e_{Y_{i}}|Y_{i}\in{\bf e}\}$ provided by the evidences has to be associated with the appropriate clusters. A valuation $e_{Y_{i}}$ for a variable $Y_{i}$ can be associated with a cluster having $Y_{i}$ alone. In our example, if the variable O1 is evidenced, then the corresponding valuation $e_{O1}$ can be associated with cluster {\bf C1}. While finding the probability distribution of a variable $Y_{i}$, the inference mechanism (as explained before) with an evidence set ${\bf e}$ will give the probability $P(Y_{i},{\bf e})$ instead of $P(Y_{i})$. From $P(Y_{i},{\bf e})$, $P({\bf e})$ is calculated as, $P({\bf e})=\sum_{Y_{i}}P(Y_{i},{\bf e})$. Calculation of the probability of evidence $P({\bf e})$ is crucial for MAP calculation.

The MAP probabilities $MAP({\bf i}_{inter},{\bf o})$ are calculated by performing inference on the binary join tree with evidences ${\bf i}_{inter}$ and ${\bf o}$. Let us say that we have an evidence set ${\bf e}=\{{\bf i}_{inter},{\bf o}\}$, then $MAP({\bf i}_{inter},{\bf o})=P({\bf e})$. For a given partial instantiation ${\bf i}_{inter}$, $MAP({\bf i}_{inter},{\bf o})$ is calculated by maximizing over the MAP variables which are not evidenced. This calculation can be done by modifying the message passing scheme to accommodate maximization over unevidenced MAP variables. So for MAP calculation, the marginalization operation involves both maximization and summation functions. The maximization is performed over the unevidenced MAP variables in {\bf I} and the summation is performed over all the other variables in {\bf X} and {\bf O}. For MAP, a message passed from cluster $b$ to cluster $c$ is calculated as,
\begin{equation}
M_{b\rightarrow c} = \max_{\{{\bf I}_{b}\}\in\{{\bf B}\setminus {\bf C}\}} \sum_{\{{\bf X}_{b}\cup{\bf O}_{b}\}\in\{{\bf B}\setminus {\bf C}\}} \phi_{b} \prod_{a\neq c} M_{a\rightarrow b}
\label{map-eqn}
\end{equation}
where ${\bf I}_{b}\subseteq{\bf I}\setminus{\bf I}_{inter}$, ${\bf X}_{b}\subseteq{\bf X}$, ${\bf O}_{b}\subseteq{\bf O}$ and $\{{\bf I}_{b},{\bf X}_{b},{\bf O}_{b}\}\in{\bf B}$. 

\begin{figure}
\begin{center}
\epsfxsize 250pt
\epsffile{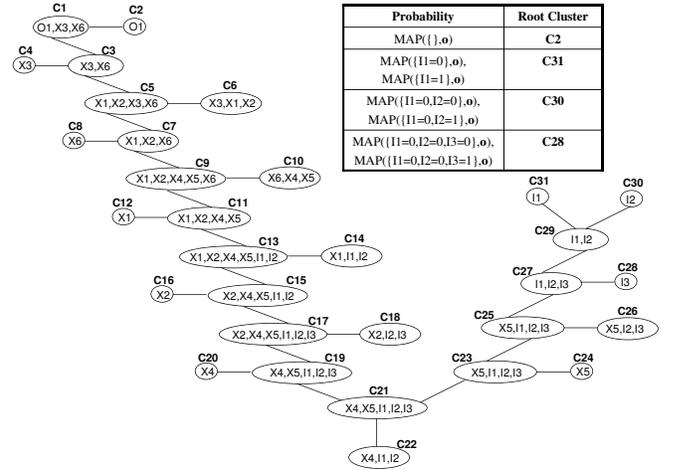}
\end{center}
\vspace*{-0.15in}
\caption{Binary join tree for the probabilistic error model in Fig.~\ref{model_fig}(c).}
\label{jointreeexample}
\end{figure}

\begin{figure*}
\begin{center}
\epsfxsize 350pt
\epsffile{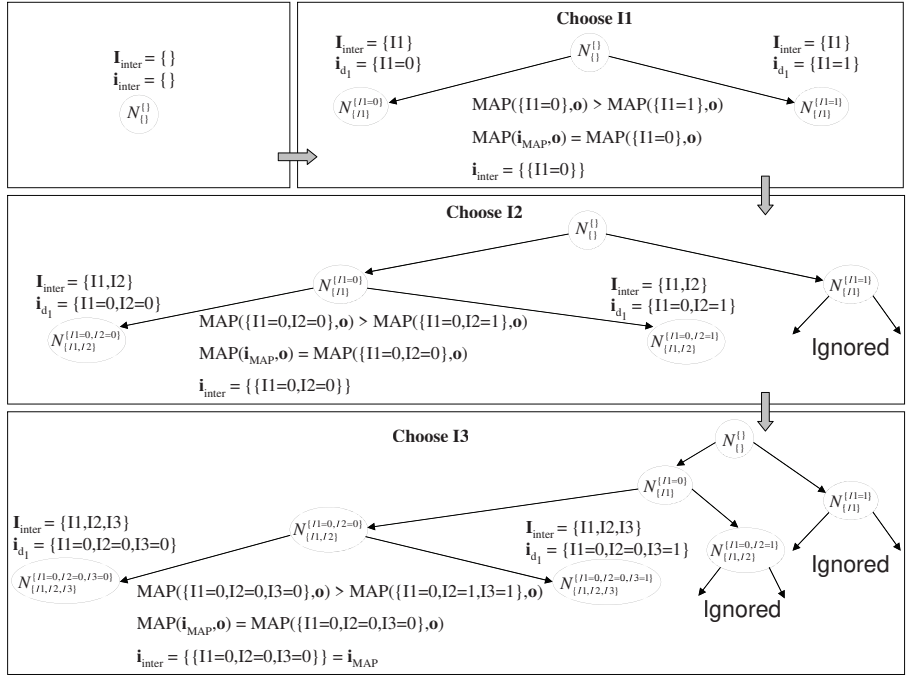}
\end{center}
\vspace*{-0.2in}
\caption{Search process for MAP computation.}
\vspace*{-0.1in}
\label{searchprocess}
\end{figure*}

Here the most important aspect is that the maximization and summation operators in Eq.~\ref{map-eqn} are non-commutative. 
\begin{equation}
  [\sum_{{\bf X}}\max_{{\bf I}}P]({\bf y}) \geq [\max_{{\bf I}}\sum_{{\bf X}}P]({\bf y})
\label{non-comm}
\end{equation}
So during message passing in the binary join tree, the {\it valid order of the marginalizing variables} or the {\it valid variable elimination order} should have the summation variables in {\bf X} and {\bf O} before the maximization variables in {\bf I}. A message pass through an invalid variable elimination order can result in a bad upper bound that is stuck at a local maxima and it eventually results in the elimination of some probable instantiations of the MAP variables {\bf I} during the search process. But an invalid elimination order can provide us an initial upper bound of the MAP probability to start with. The closer the invalid variable elimination order to the valid one, the tighter will be the upper bound. In the binary join tree, any cluster can be chosen as root to get this initial upper bound. For example, in Fig.~\ref{messagepassing}(b) choosing cluster {\bf C1} as root results in an invalid variable elimination order I1,I2,X1,X2 and message pass towards this root can give the initial upper bound. Also it is essential to use a valid variable elimination order during the construction of the binary join tree so that there is at least one path that can provide a good upper bound. 

Fig.~\ref{jointreeexample} gives the corresponding binary join tree, for the probabilistic error model given in Fig.~\ref{model_fig}(c), constructed with a valid variable elimination order (O1,X3,X6,X1,X2,X4,X5,I3,I2,I1). In this model, there are three MAP variables I1,I2,I3. The MAP hypothesis on this model results in ${\bf i}_{MAP} = \{I1=0,I2=0,I3=0\}$. 

The initial upper bound $MAP(\{\},{\bf o})$ is calculated by choosing cluster {\bf C2} as root and passing messages towards {\bf C2}. As specified earlier this upper bound can be calculated with any cluster as root. With {\bf C2} as root, an upper bound will most certainly be obtained since the variable elimination order (I3,I2,I1,X4,X5,X1,X2,X3,X6) is an invalid one. But since the maximization variables are at the very beginning of the order, having {\bf C2} as root will yield a looser upper bound. Instead, if {\bf C16} is chosen as root, the elimination order (O1,X3,X6,X1,I3,X4,X5,I2,I1) will be closer to a valid order. So a much tighter upper bound can be achieved. To calculate an intermediate upper bound $MAP({\bf i}_{inter},{\bf o})$, the MAP variable $I_{i}$ newly added to form ${\bf i}_{inter}$ is recognized and the cluster having the variable $I_{i}$ alone is selected as root. By doing this a valid elimination order and proper upper bound can be achieved. For example, to calculate the intermediate upper bound $MAP(\{I1=0\},{\bf o})$ where the instantiation $\{I1=0\}$ is newly added to the initially empty set ${\bf i}_{inter}$, a valid elimination order should have the maximization variables I2,I3 at the end. To achieve this, cluster ${\bf C31}$ is chosen as root thereby yielding a valid elimination order (O1,X3,X6,X1,X2,X4,X5,I3,I2). 

\subsection{Calculation of the exact MAP solution}
\label{final_map}

The calculation of the exact MAP solution $MAP({\bf i}_{MAP},{\bf o})$ can be explained as follows,
\begin{enumerate}
\item To start with we have the following,\\
${\bf I}_{inter} \rightarrow$ subset of MAP variables ${\bf I}$. Initially empty.\\
${\bf i}_{inter} \rightarrow$ partial instantiation set of MAP variables ${\bf I}_{inter}$. Initially empty.\\
${\bf i}_{d_{1}},{\bf i}_{d_{2}} \rightarrow$ partial instantiation sets used to store ${\bf i}_{inter}$. Initially empty.\\
${\bf i}_{MAP} \rightarrow$ MAP instantiation. At first, ${\bf i}_{MAP}={\bf i}_{init}$, where ${\bf i}_{init}$ is calculated by {\it sequentially} initializing the MAP variables to a particular instantiation and performing local {\it taboo search} around the neighbors of that instantiation~\cite{hill_climb}. Since this method is out of the scope of this paper, we are not explaining it in detail.\\
$MAP({\bf i}_{MAP},{\bf o}) \rightarrow$ MAP probability. Initially $MAP({\bf i}_{MAP},{\bf o})=MAP({\bf i}_{init},{\bf o})$ calculated by inferencing the probabilistic error model.\\
$v(I_{i}) \rightarrow$ number of values or states that can be assigned to a variable $I_{i}$. Since we are dealing with digital signals, $v(I_{i}) = 2$ for all $i$.\\

\item 

\begin{algorithmic}[1]
\STATE Calculate $MAP({\bf i}_{inter},{\bf o})$. /*{\it This is the initial upper bound of MAP probability.}*/
\IF{$MAP({\bf i}_{inter},{\bf o}) \geq MAP({\bf i}_{MAP},{\bf o})$}
	\STATE $MAP({\bf i}_{MAP},{\bf o}) = MAP({\bf i}_{inter},{\bf o})$
\ELSE
	\STATE $MAP({\bf i}_{MAP},{\bf o}) = MAP({\bf i}_{MAP},{\bf o})$	
	\STATE ${\bf i}_{MAP}={\bf i}_{MAP}$
\ENDIF
\WHILE{$|{\bf I}| > 0$}
	\STATE Choose a variable $I_{i}\in{\bf I}$.
	\STATE ${\bf I}_{inter}={\bf I}_{inter}\cup\{I_{i}\}$.
	\WHILE{$v(I_{i}) > 0$}
		\STATE Choose a value $i_{v(I_{i})}$ of $I_{i}$
		\STATE ${\bf i}_{d_{1}}={\bf i}_{inter}\cup\{I_{i}=i_{v(I_{i})}\}$.
		\STATE Calculate $MAP({\bf i}_{d_{1}},{\bf o})$ from binary join tree.
		\IF{$MAP({\bf i}_{d_{1}},{\bf o}) \geq MAP({\bf i}_{MAP},{\bf o})$}
			\STATE $MAP({\bf i}_{MAP},{\bf o}) = MAP({\bf i}_{d_{1}},{\bf o})$
			\STATE ${\bf i}_{d_{2}}={\bf i}_{d_{1}}$
		\ELSE
			\STATE $MAP({\bf i}_{MAP},{\bf o}) = MAP({\bf i}_{MAP},{\bf o})$
		\ENDIF	
		\STATE $v(I_{i}) = v(I_{i})-1$
	\ENDWHILE	
	\STATE ${\bf i}_{inter}={\bf i}_{d_{2}}$
	\IF{$|{\bf i}_{inter}|=0$}
		\STATE goto line 29
	\ENDIF	
	\STATE ${\bf I}={\bf I}-\{I_{i}\}$ 
\ENDWHILE
\IF{$|{\bf i}_{inter}|=0$}
	\STATE ${\bf i}_{MAP}={\bf i}_{MAP}$
\ELSE
	\STATE ${\bf i}_{MAP}={\bf i}_{inter}$	
\ENDIF
\end{algorithmic}

\end{enumerate}

The pruning of the search process is handled in lines 11-23. After choosing a MAP variable $I_{i}$, the partial instantiation set ${\bf i}_{inter}$ is updated by adding the best instantiation $I_{i}=i_{v(I_{i})}$ thereby ignoring the other instantiations of $I_{i}$. This can be seen in Fig.~\ref{searchprocess} which  illustrates the search process for MAP computation using the probabilistic error model given in Fig.~\ref{model_fig}(c) as example.

\subsection{Calculating the maximum output error probability}
\label{maxerror}

According to our error model, the MAP variables represent the primary input signals of the underlying digital logic circuit. So after MAP hypothesis, we will have the input vector which has the highest probability to give an error on the output. The random variables {\bf I} that represent the primary input signals are then instantiated with ${\bf i}_{MAP}$ and inferenced. So the evidence set for this inference calculation will be ${\bf e}=\{{\bf i}_{MAP}\}$. The output error probability is obtained by observing the probability distributions of the comparator logic variables {\bf O}. After inference, the probability distribution $P(O_{i},{\bf e})$ will be obtained. From this $P(O_{i}|{\bf e})$ can be obtained as, $P(O_{i}|{\bf e})=\frac{P(O_{i},{\bf e})}{P({\bf e})}=\frac{P(O_{i},{\bf e})}{\sum_{O_{i}} P(O_{i},{\bf e})}$. Finally the maximum output error probability is given by, $\max_{i} P(O_{i}=1|{\bf e})$.

\subsection{Computational complexity of MAP estimate}
\label{complexity}

The time complexity of MAP depends on that of the depth first branch and bound search on the {\it input instantiation search tree} and also on that of {\it inference in binary join tree}. The former depends on the number of MAP variables and the number of states assigned to each variable. In our case each variable is assigned two states and so the time complexity can be given as $O(2^{k})$ where $k$ is the number of MAP variables. This is the worst case time complexity assuming that the search tree is not pruned. If the search tree is pruned, then the time complexity will be $< O(2^{k})$. 

The time complexity of inference in the binary join tree depends on the number of cliques $q$ and the size $Z$ of the biggest clique. It can be represented as $q.2^{Z}$ and the worst case time complexity can be given as $O(2^{Z})$. In any given probabilistic model with $N$ variables, representing a joint probability $P(x_{1},\cdots x_{N})$, the corresponding jointree will have $Z < N$ always~\cite{hugin}. Also depending on the underlying circuit structure, the jointree of the corresponding probabilistic error model can have $Z << N$ or $Z$ close to $N$, which in turn determines the time complexity.

Since for every pass in the search tree inference has to be performed in the join tree to get the upper bound of MAP probability, the worst case time complexity for MAP can be given as $O(2^{k + Z})$. The space complexity of MAP depends on the number of MAP variables for the search tree and on the number of variables $N$ in the probabilistic error model and the size of the largest clique. It can be given by $2^{k} + N.2^{Z}$.  

\section{Experimental Results}
\label{results}

\begin{figure}
\vspace*{-0.4in}
\begin{center} 
\epsfxsize 320pt 
{\epsffile{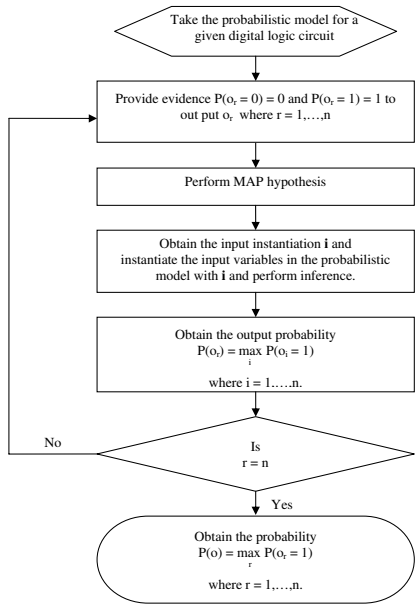}} 
\end{center} 
\vspace*{-2.2in}
\caption{Flow chart describing the experimental setup and process} 
\label{chart} 
\end{figure}

The experiments are performed on ISCAS85 and MCNC benchmark circuits. 
The computing device used is a Sun server with 8 CPUs where each CPU consists of 1.5GHz UltraSPARC IV processor with at least 32GB of RAM. 

\subsection{Experimental procedure for calculating maximum output error probability}

Our main goal is to provide the maximum output error probabilities
for different gate error probabilities $\varepsilon$. To get the
maximum output error probabilities every output signal of a circuit
has to be examined through MAP estimation, which is performed through algorithms provided in~\cite{samiam}. The experimental procedure is illustrated as a flow chart in Fig.~\ref{chart}. The steps are as follows,

\begin{enumerate}

\item First, an evidence has to be provided to one of the comparator output signal variables in set {\bf O} such that $P(O_{i}=0)=0$ and $P(O_{i}=1)=1$. Recall that these variables have a probability distribution based on XOR logic and so giving evidence like this is similar to forcing the output to be wrong. 

\item The comparator outputs are evidenced individually and the corresponding input instantiations {\bf i} are obtained by performing MAP. 

\item Then the primary input variables in the probabilistic error model are instantiated with each instantiation {\bf i} and inferenced to get the output probabilities. 

\item $P(O_{i}=1)$ is noted from all the comparator outputs for each {\bf i} and the maximum value gives the maximum output error probability. 

\item The entire operation is repeated for different $\varepsilon$ values.

\end{enumerate}

\begin{figure*}
\begin{center}
\begin{tabular}{c c c c}
\epsfxsize 0.45\columnwidth
\epsffile{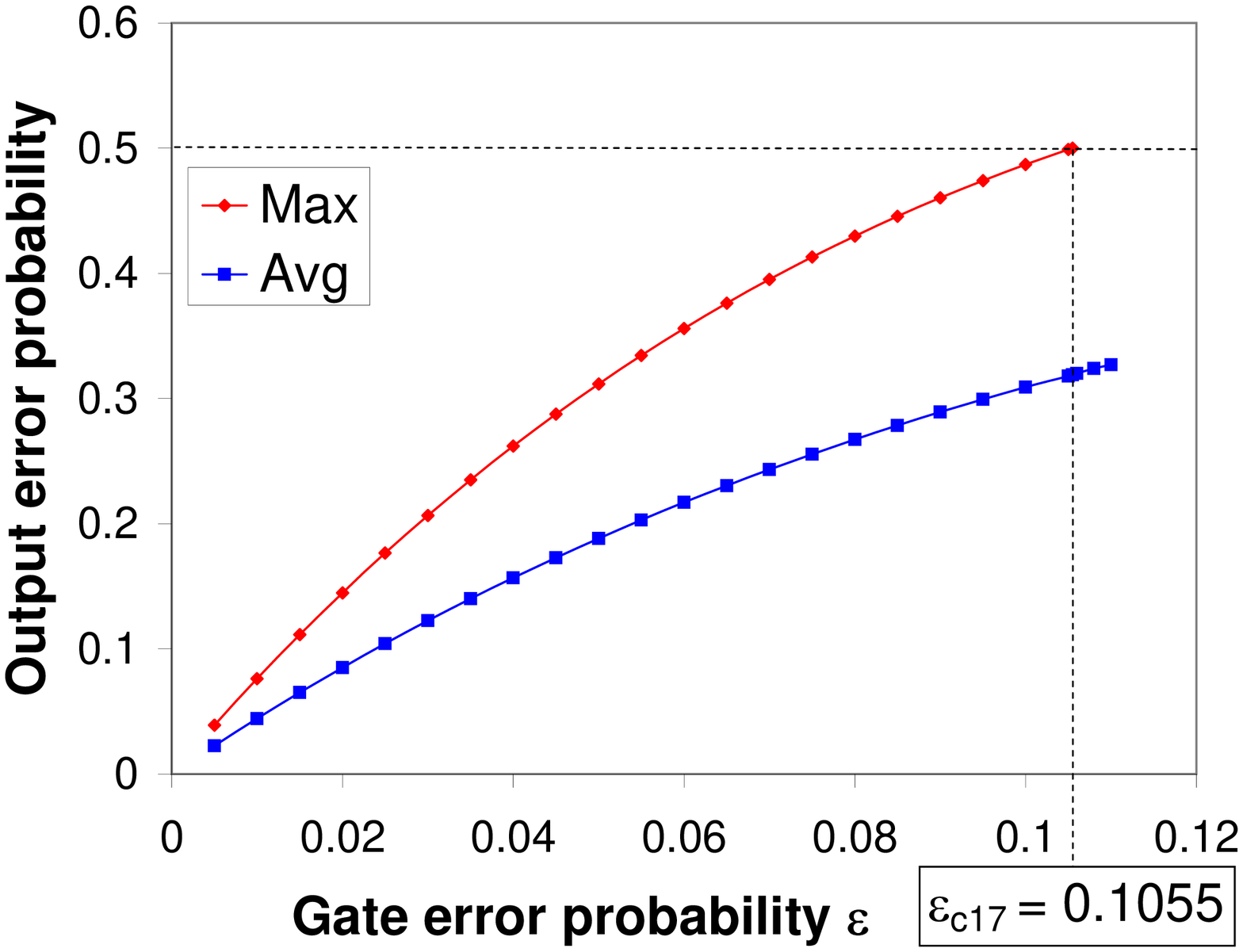}
&
\epsfxsize 0.45\columnwidth
\epsffile{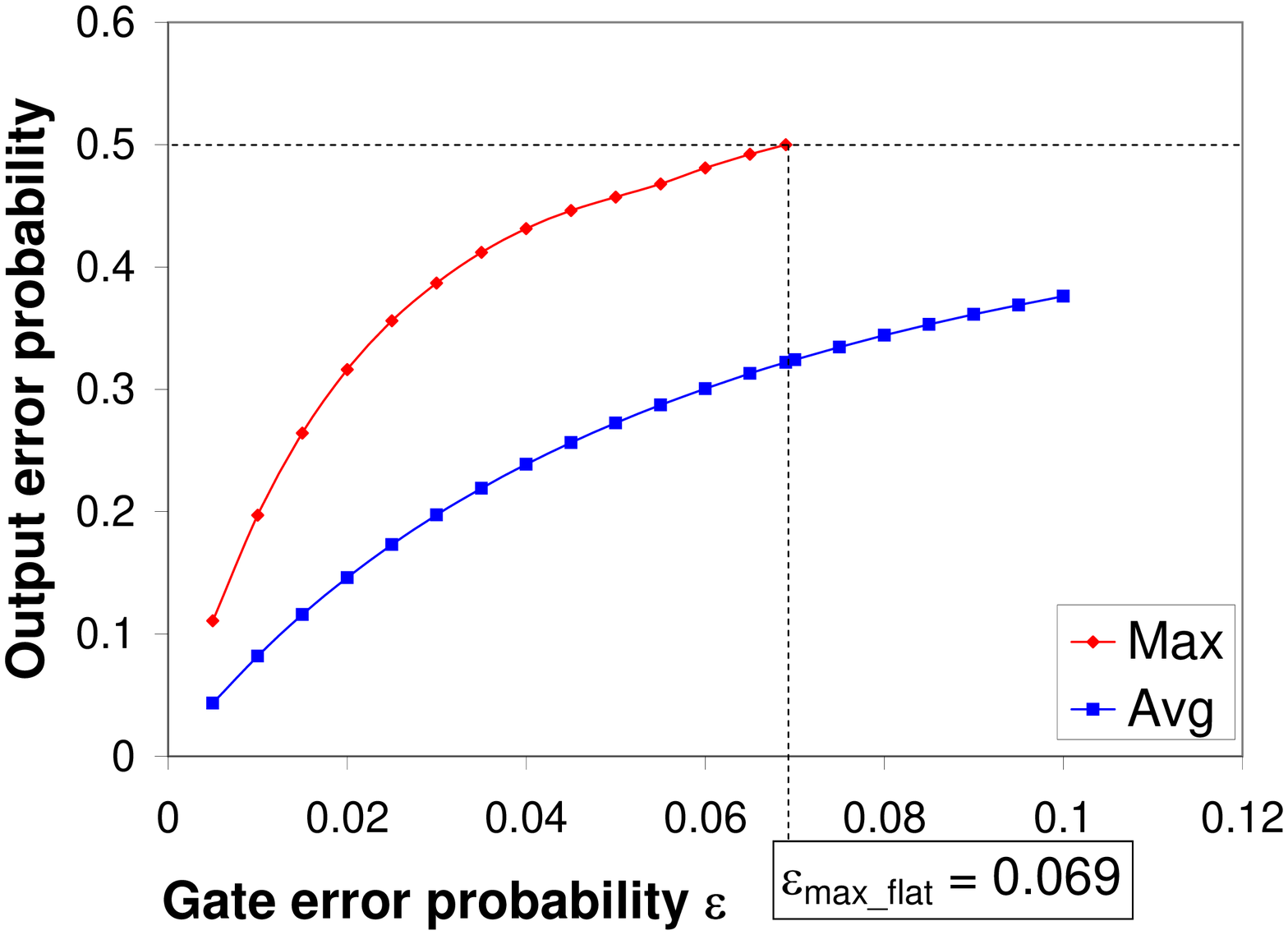}
&
\epsfxsize 0.45\columnwidth
\epsffile{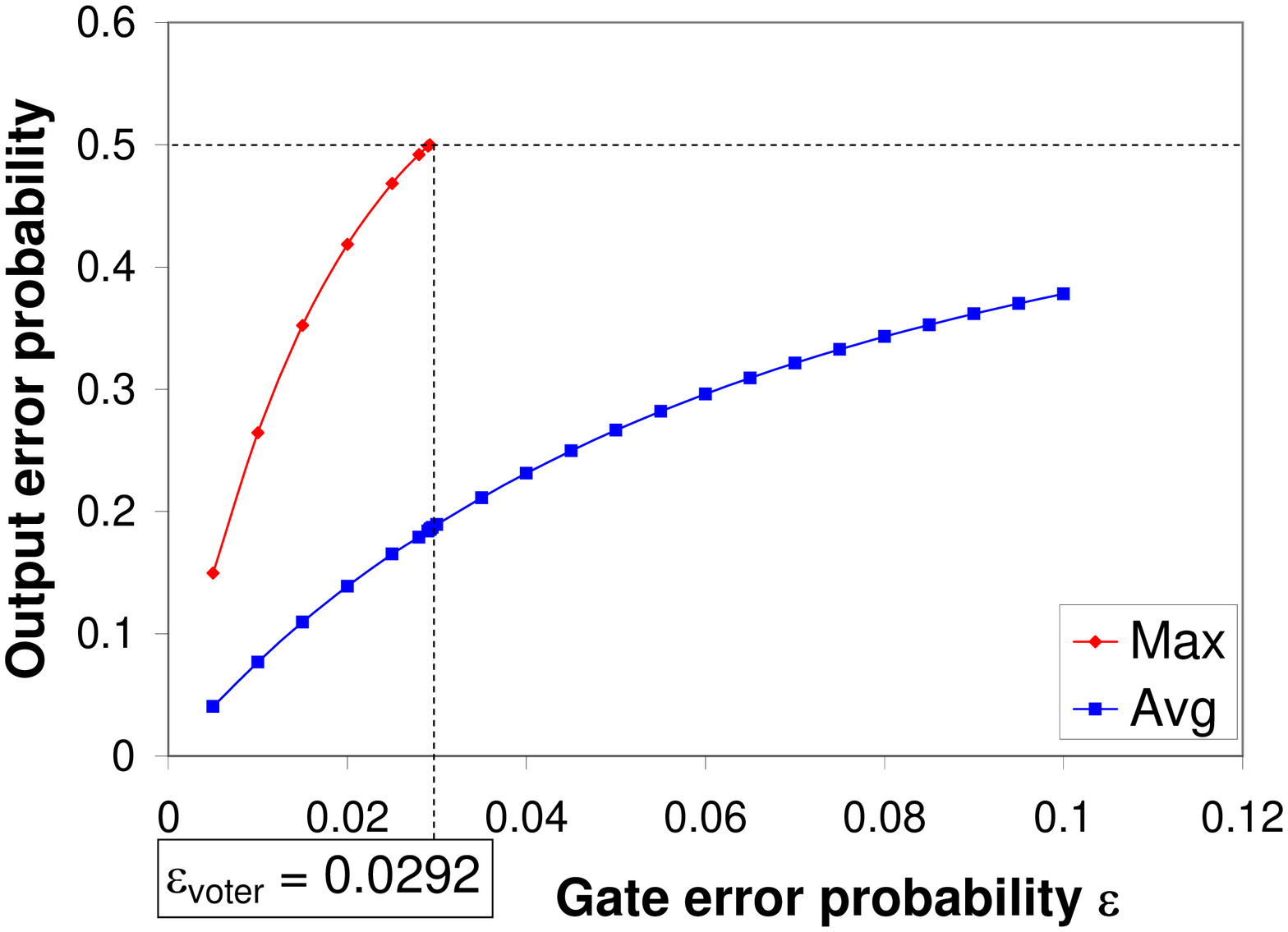}
&
\epsfxsize 0.45\columnwidth
\epsffile{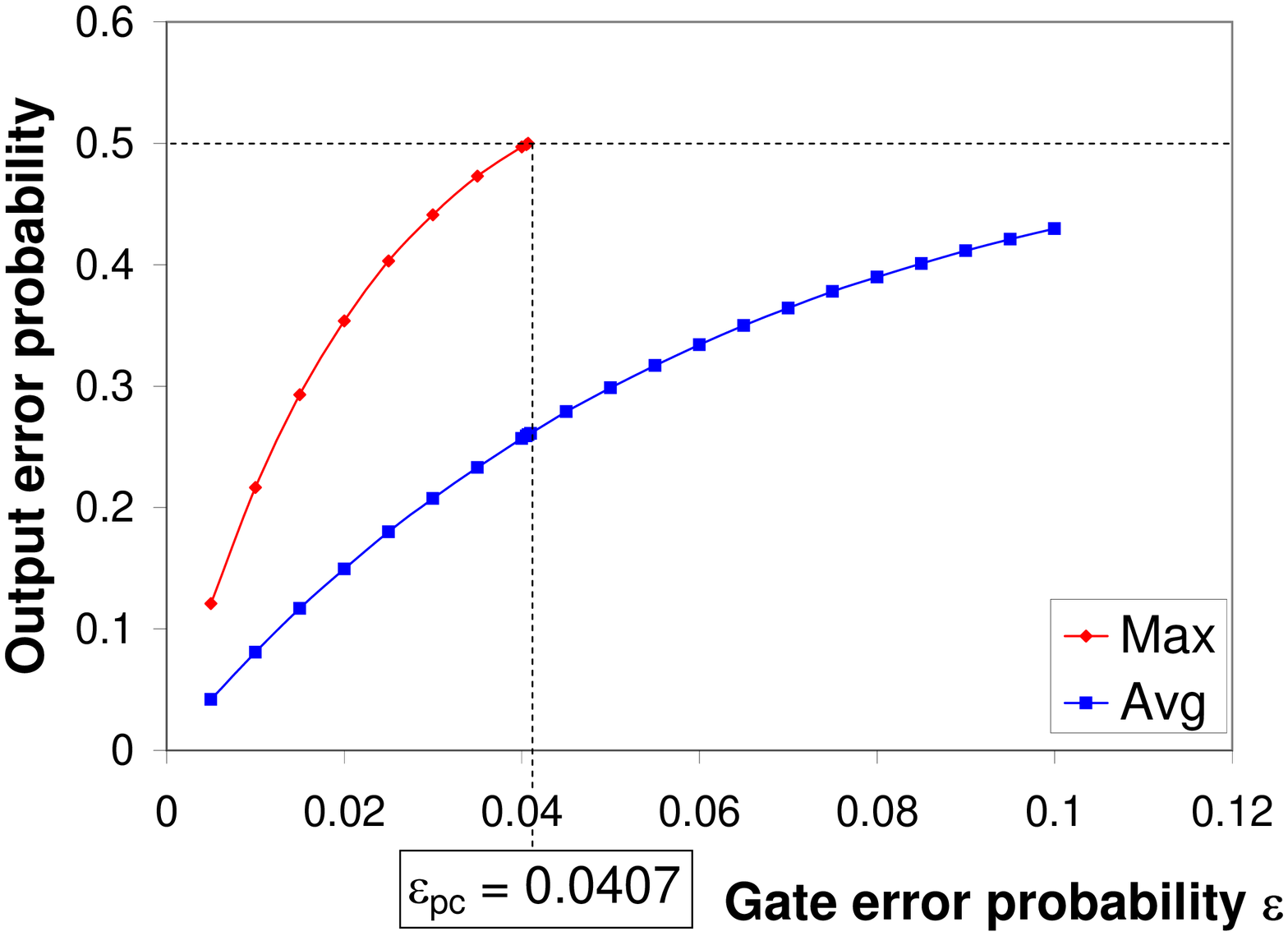}
\\ (a) & (b) & (c) & (d) \\
\epsfxsize 0.45\columnwidth
\epsffile{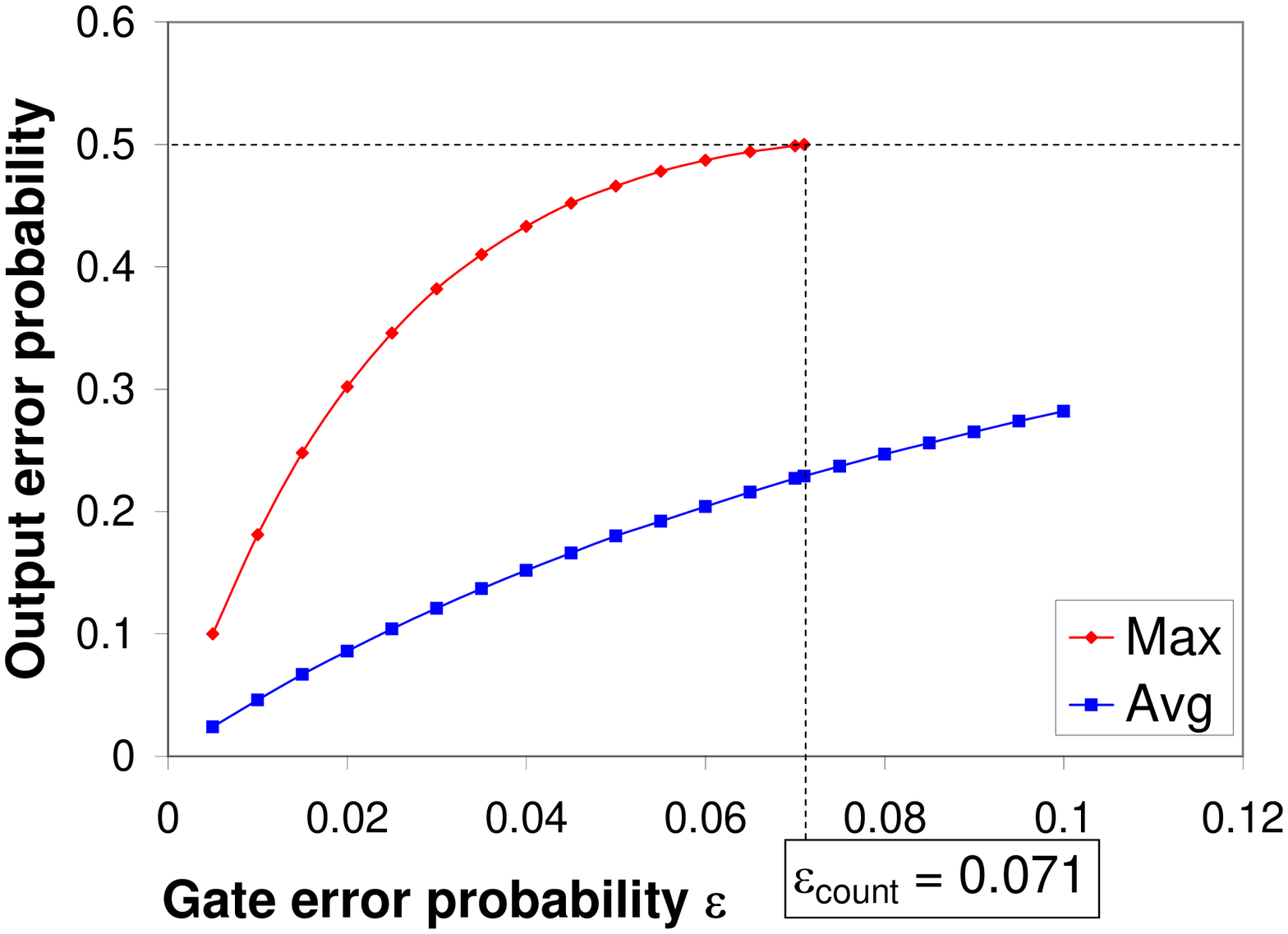}
&
\epsfxsize 0.45\columnwidth
\epsffile{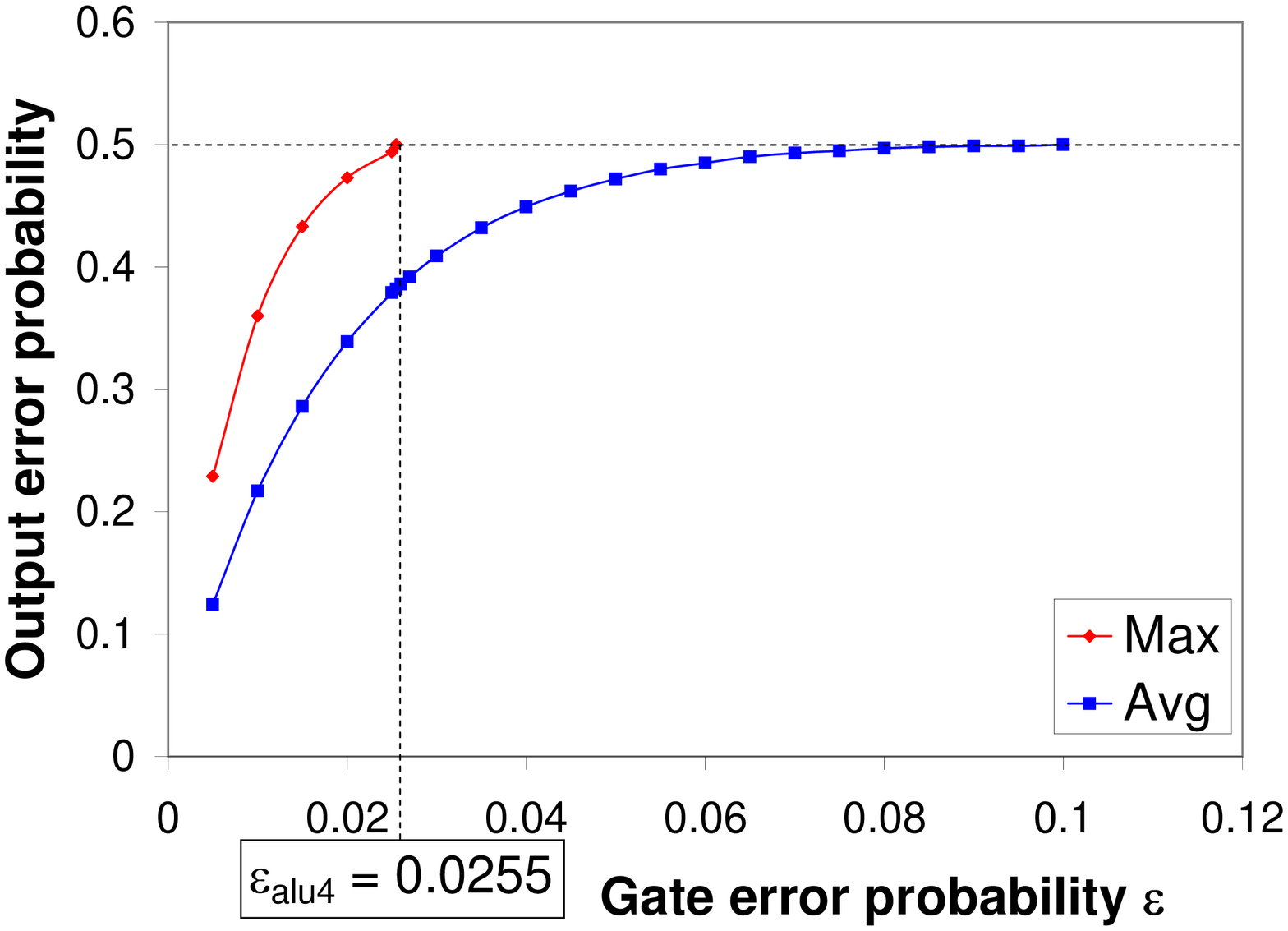}
&
\epsfxsize 0.45\columnwidth
\epsffile{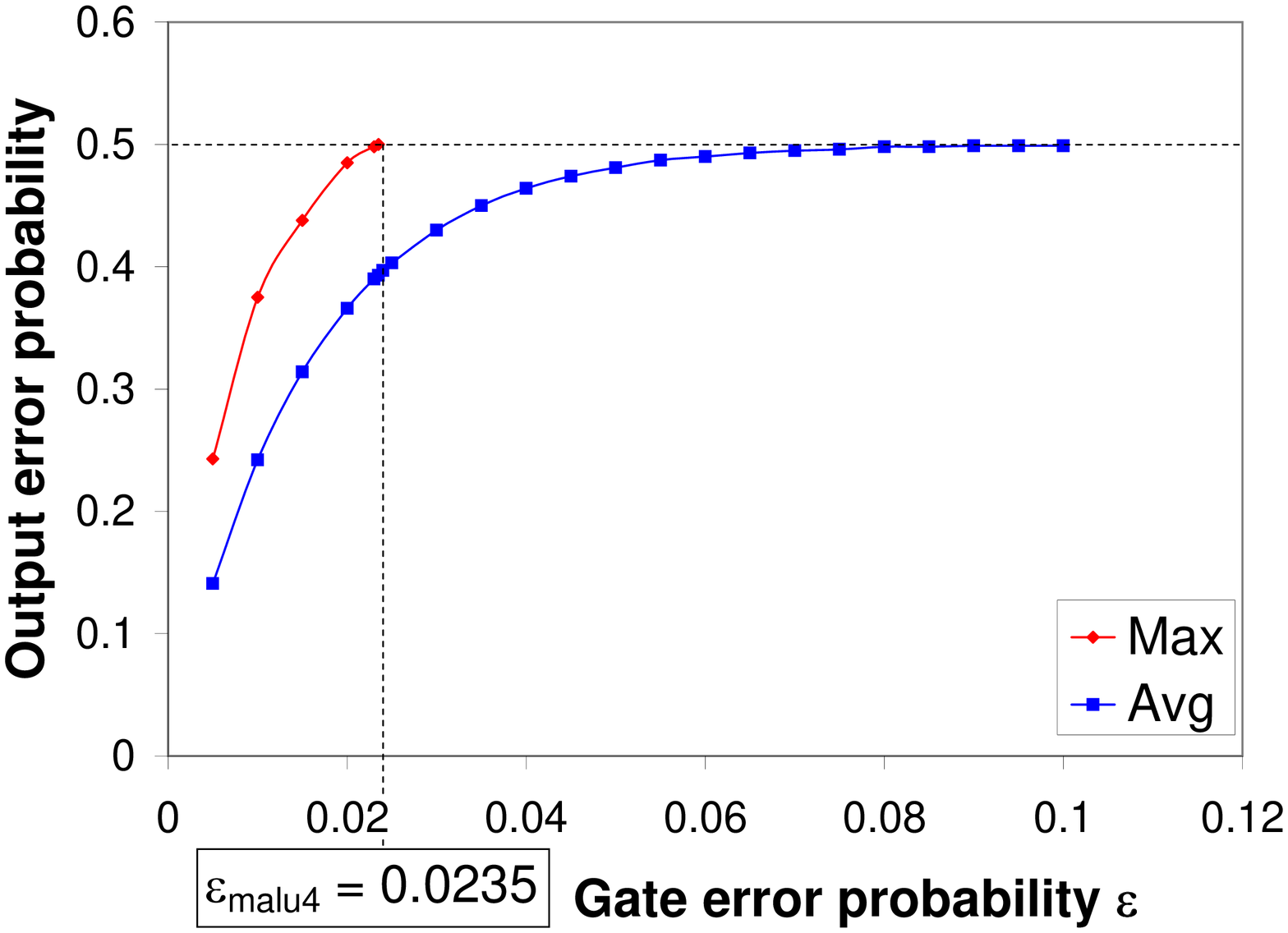}
&
\\ (e) & (f) & (g) &
\end{tabular}
\end{center}
\caption{Circuit-specific error bound for (a) $c17$, (b) $max\_flat$, (c) $voter$, (d) $pc$, (e) $count$, (f) $alu4$, (g) $malu4$. The figures also show the comparison between maximum and average output error probabilities, that indicates the importance of using maximum output error probability to achieve a tighter error bound.}
\label{bounds}
\end{figure*}

\subsection{Worst-case Input Vectors} 

\begin{table} 
{\scriptsize
\caption{Worst-case input vectors from MAP} 
\label{input_comb} 
\begin{center} 
\begin{tabular}{||c|c|c|c||} \hline  
Circuits&  No. of & Input vector & Gate error \\
        &  Inputs &              &  probability $\varepsilon$\\ \hline 
c17     &      5          &  01111 & 0.005 - 0.2 \\ \hline 
max\_flat&     8          & 00010011 & 0.005 - 0.025\\          
&                & 11101000 & 0.03 - 0.05\\
         &                & 11110001 & 0.055 - 0.2\\ \hline
voter   &      12         & 000100110110 & 0.01 - 0.19\\
          &                & 111011100010 & 0.2\\ \hline
\end{tabular}  
\end{center}  
}
\end{table} 

{\bf Table~\ref{input_comb}} gives the worst-case input vectors got
from MAP i.e., the input vectors that gives maximum output error probability. The notable results are as follows,

\begin{itemize}

\item In $max\_flat$ and $voter$ the worst-case input vectors from MAP
changes with $\varepsilon$, while in $c17$ it does not change. 

\item In the range \{0.005-0.2\} for $\varepsilon$, $max\_flat$ has three different worst-case input vectors while $voter$ has two. 

\item It implies that these worst-case input vectors not only depend on the circuit structure but could dynamically change with $\varepsilon$. This
could be of concern for designers as the worst-case inputs might change
after gate error probabilities reduce due to error mitigation schemes.
Hence, explicit MAP computation would be necessary to judge the
maximum error probabilities and worst-case vectors after every redundancy schemes are applied.

\end{itemize}  

\subsection{Circuit-specific error bounds for fault-tolerant computation}

The error bound for a circuit can be obtained by calculating the gate error probability $\varepsilon$ that drives the output error probability of at least one output to a hard bound beyond which the output does not depend on the input signals or the circuit structure. When the output error probability reaches $0.5 (50\%)$, it essentially means that the output signal behaves as a non-functional random number generator for at least one input vector and so $0.5$ can be treated as a hard bound.

{\bf Fig.~\ref{bounds}} gives the error bounds for various benchmark circuits. It also shows the comparison between maximum and average output error probabilities with reference to the change in gate error probability $\varepsilon$. These graphs are obtained by performing the experiment for different $\varepsilon$ values ranging from $0.005$ to $0.1$. The average error probabilities are obtained from our previous work by Rejimon et.al~\cite{thara}. The notable results are as follows,

\begin{itemize}

\item The $c17$ circuit consists of 6 NAND gates. The error bound for each NAND gate in $c17$ is $\varepsilon = 0.1055$, which is greater than the conventional error bound for NAND gate, which is $0.08856$~\cite{evans, jose02}. The error bound of the same NAND gate in $voter$ circuit (contains 10 NAND gates, 16 NOT gates, 8 NOR gates, 15 OR gates and 10 AND gates) is $\varepsilon = 0.0292$, which is lesser than the conventional error bound. This indicates that the error bound for an individual {\it NAND gate placed in a circuit} can be dependent on the circuit structure. The same can be true for all other logics.

\item The maximum output error probabilities are much larger than average output error probabilities, thereby reaching the hard bound for comparatively lower values of $\varepsilon$, making them a very crucial design parameter to achieve tighter error bounds. Only for $alu4$ and $malu4$, the average output error probability reaches the hard bound within $\varepsilon = 0.1 (\varepsilon = 0.095 for ~alu4, \varepsilon = 0.08 for ~malu4)$, while the maximum output error probabilities for these circuits reach the hard bound for far lesser gate error probabilities ($\varepsilon = 0.0255 for ~alu4, \varepsilon = 0.0235 for ~malu4$). 
\item While the error bounds for all the circuits, except $c17$, are less than $0.08 (8\%)$, the error bounds for circuits like $voter$, $alu4$ and $malu4$ are even less than $0.03 (3\%)$ making them highly vulnerable to errors.

\end{itemize}

\begin{table} 
{\scriptsize
\caption{Run times for MAP computation} 
\label{result_table} 
\begin{center} 
\begin{tabular}{||c|c|c|c||} \hline  
\multirow{2}{*}{Circuit} & No. of & No. of &	
\multirow{2}{*}{Time} \\ 
& Inputs & Gates & \\ \hline
c17 & 5 & 6 & 0.047s\\ \hline
max\_flat & 8 & 29 & 0.110s\\ \hline
voter & 12 & 59 & 0.641s\\ \hline
pc & 27 & 103 & 225.297s\\ \hline
count & 35 & 144 & 36.610s\\ \hline
alu4 & 14 & 63 & 58.626s\\ \hline
malu4 & 14 & 92 & 588.702s\\ \hline
\end{tabular}  
\end{center}  
}
\end{table} 

{\bf Table~\ref{result_table}} tabulates the run time for MAP computation. The run time does not change significantly for different $\varepsilon$ values and so we provide only one run time which corresponds to all $\varepsilon$ values. This is expected as MAP complexity (discussed in Sec.~\ref{complexity}) is determined by number of inputs, and number of variables in the largest clique which in turn depends on the circuit complexity. It has to be noted that, even though $pc$ has less number of inputs than $count$, it takes much more time to perform MAP estimate due to its complex circuit structure. 

\subsection{Validation using HSpice simulator} 
 
\begin{figure*} 
\begin{center} 
\begin{tabular}{c c c}
\epsfxsize 0.66\columnwidth 
{\epsffile{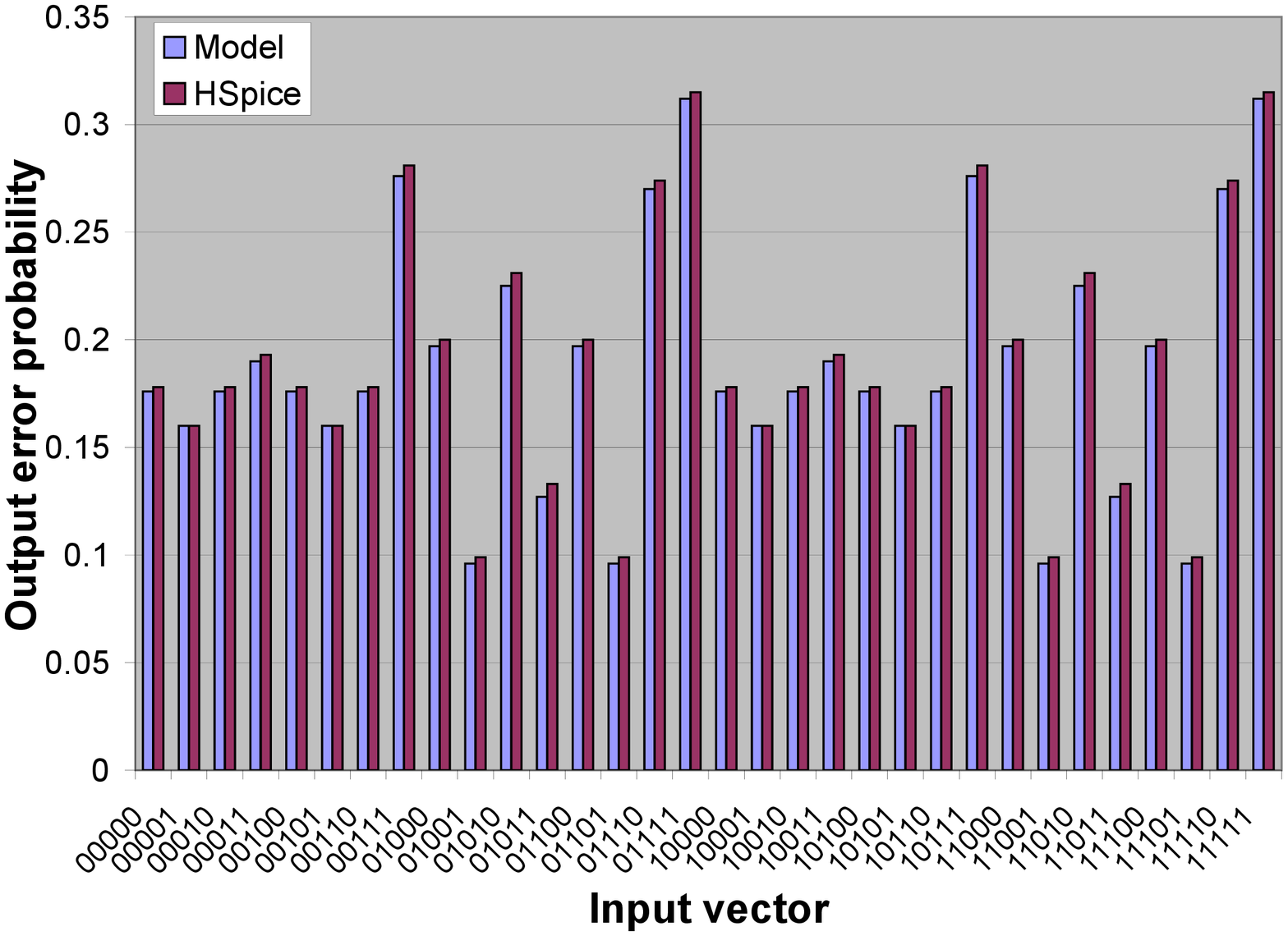}} 
&
\epsfxsize 0.66\columnwidth
{\epsffile{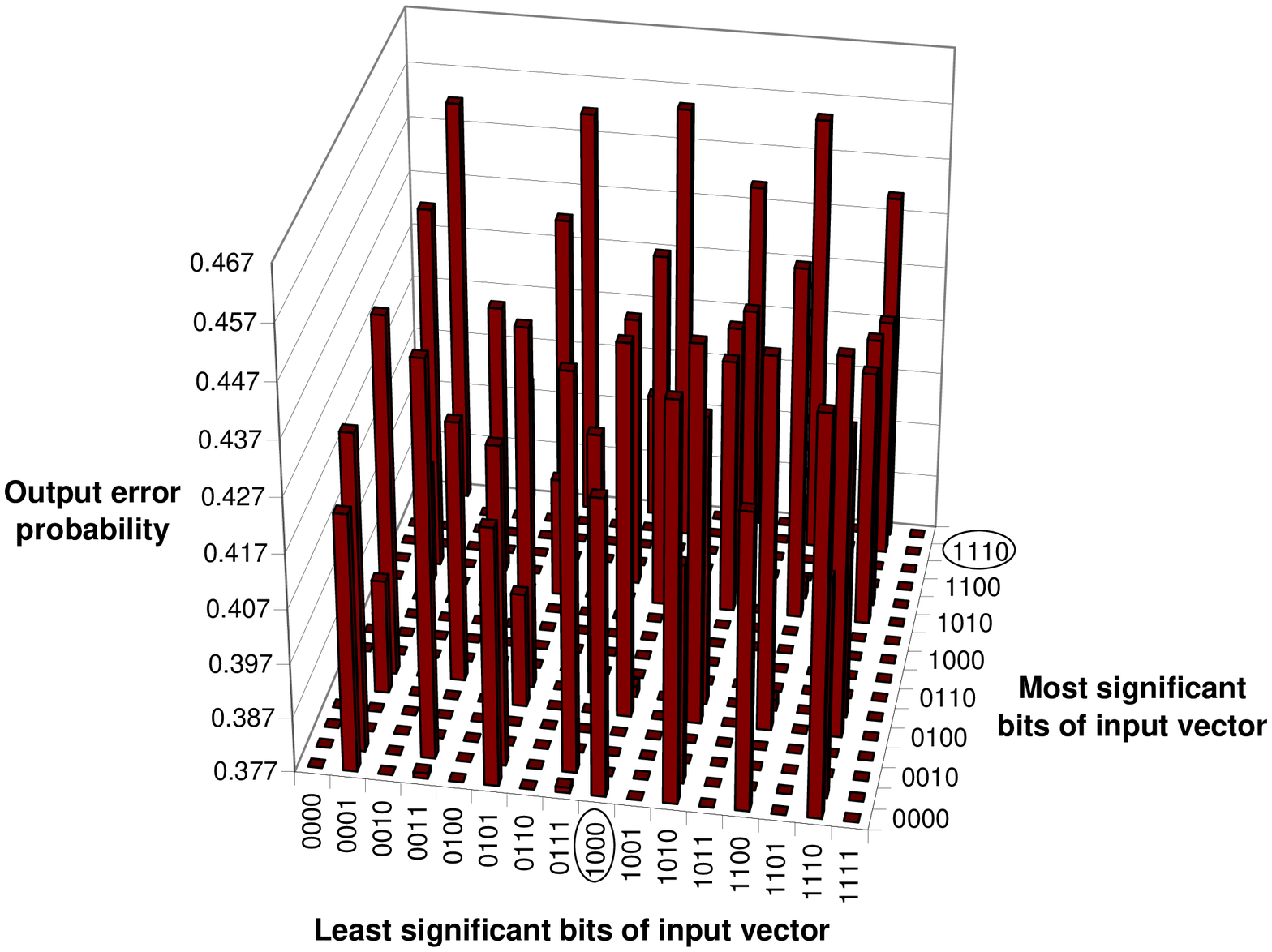}} 
&
\epsfxsize 0.66\columnwidth
{\epsffile{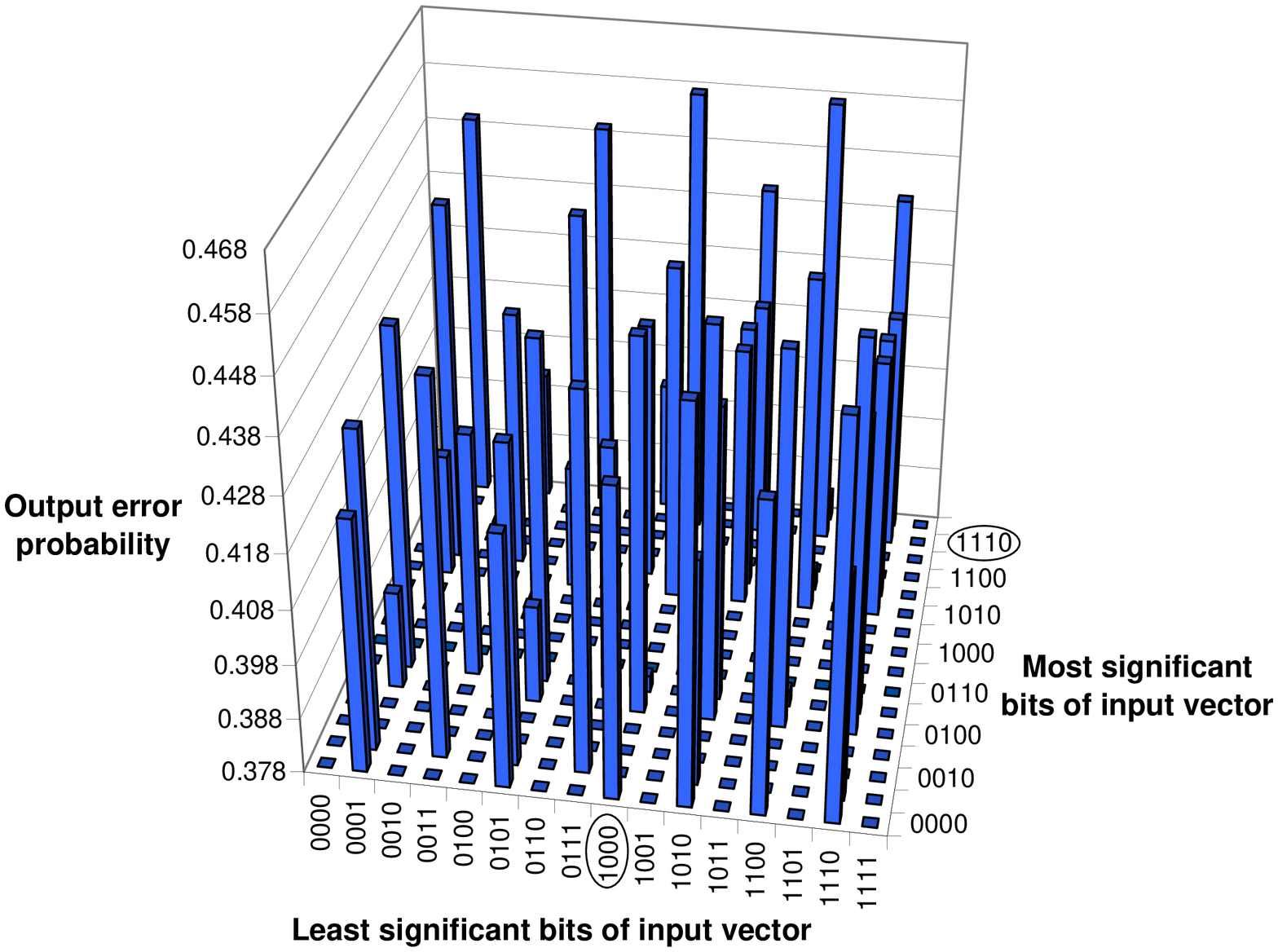}} 
\\ (a) & (b) & (c)
\end{tabular}
\end{center} 
\caption{(a) Output error probabilities for the entire input vector space with gate error probability $\varepsilon = 0.05$ for $c17$. (b) Output error probabilities $\geq (\mu + \sigma)$, calculated from probabilistic error model, with gate error probability $\varepsilon = 0.05$ for $max\_flat$. (c) Output error probabilities $\geq (\mu + \sigma)$, calculated from HSpice, with gate error probability $\varepsilon = 0.05$ for $max\_flat$ }
\label{maxerrorvector} 
\end{figure*}

\begin{table}
\caption{Comparison between Maximum error probabilities achieved from the proposed model and the HSpice simulator at $\varepsilon=0.05$ }
\label{comp_table} 
\begin{center} 
\begin{tabular}{||c|c|c|c||} \hline  
Circuit & Model & HSpice & \% diff over HSpice \\ \hline  
c17     & 0.312	& 0.315	& 0.95 \\ \hline
max\_flat & 0.457 & 0.460 & 0.65 \\ \hline
voter	& 0.573	& 0.570	& 0.53 \\ \hline
pc	& 0.533	& 0.536	& 0.56 \\ \hline
count	& 0.492	& 0.486	& 1.23 \\ \hline
alu4	& 0.517	& 0.523	& 1.15 \\ \hline
malu4	& 0.587	& 0.594	& 1.18 \\ \hline
\end{tabular}  
\end{center}  
\end{table} 

{\bf HSpice model}: Using external voltage sources error can be induced in any signal and it can be modeled using HSpice~\cite{pcmos}. In our HSpice model we have induced error, using external voltage sources, in every gate's output. Consider signal $O_{f}$ is the original error free output signal and the signal $O_{p}$ is the error prone output signal and $E$ is the {\it piecewise linear} (PWL) voltage source that induces error. The basic idea is that the signal $O_{p}$ is dependent on the signal $O_{f}$ and the voltage $E$. Any change of voltage in $E$ will be reflected in $O_{p}$. If $E = 0v$, then $O_{p} = O_{f}$, and if $E = Vdd~(supply~voltage)$, then $O_{p} \neq O_{f}$, thereby inducing error. The data points for the PWL voltage source $E$ are provided by computations on a finite automata which models the underlying error prone circuit where individual gates have a gate error probability $\varepsilon$.    

{\bf Simulation setup}: Note that, for an input vector of the given circuit, a single simulation run in HSpice is not enough to validate the results from our probabilistic model. Also the circuit has to be simulated for each and every possible input vectors to find out the worst-case one. For a given circuit, {\it the HSpice simulations are conducted for all possible input vectors, where for each vector the circuit is simulated for $1~million$ runs and the comparator nodes are sampled.} From this data the maximum output error probability and the corresponding worst-case input vector are obtained. 

{\bf Table~\ref{comp_table}} gives the comparison between maximum error probabilities achieved from the proposed model and the HSpice simulator at $\varepsilon=0.05$. The notable results are as follows,

\begin{itemize}

\item The simulation results from HSpice almost exactly coincides with those of our error model for all circuits.

\item The highest \% difference of our error model over HSpice is just $1.23\%$.

\end{itemize}

{\bf Fig.~\ref{maxerrorvector}}(a) gives the output error probabilities for the entire input vector space of $c17$ with gate error probability $\varepsilon = 0.05$. The notable results are as follows,

\begin{itemize}

\item It can be clearly seen that the results from {\it both} the probabilistic error model and HSpice simulations show that $01111$ gives the maximum output error probability. 

\end{itemize}

{\bf Fig.~\ref{maxerrorvector}(b) and (c)} give the output error probabilities, obtained from the probabilistic error model and HSpice respectively, for $max\_flat$ with gate error probability $\varepsilon = 0.05$. In order to show that $max\_flat$ has large number of input vectors capable of generating maximum output error, we plot output error probabilities $\geq ((\mu) + (\sigma))$, where $\mu$ is the mean of output error probabilities and $\sigma$ is the standard deviation. The notable results are as follows,

\begin{itemize}

\item It is clearly evident from Fig.~\ref{maxerrorvector}(b) that $max\_flat$ has a considerably large amount of input vectors capable of generating output error thereby making it error sensitive. Equivalent HSpice results from Fig.~\ref{maxerrorvector}(c) confirms this aspect. 

\item It is clearly evident that the results from probabilistic error model and HSpice show the same worst-case input vector, $11101000$, that is obtained through MAP hypothesis.

\end{itemize}

\subsection{Results with multiple $\varepsilon$} 

\begin{figure} 
\begin{center} 
\epsfxsize 220pt
{\epsffile{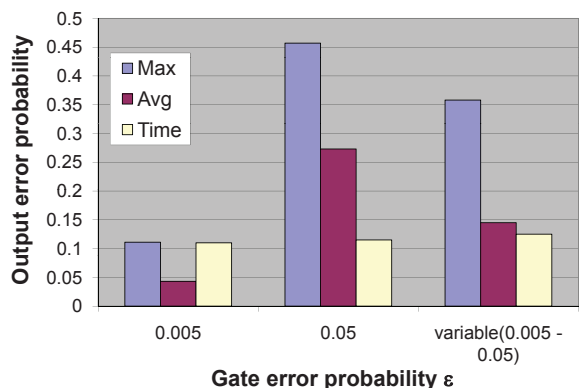}} 
\end{center} 
\vspace*{-0.2in}
\caption{Comparison between the average and maximum output error probability and run time for $\varepsilon$=0.005, $\varepsilon$=0.05 and variable $\varepsilon$ ranging for 0.005 - 0.05 for $max\_flat$}
\vspace*{-0.2in}
\label{mf_time} 
\end{figure}  

Apart from incorporating a single gate error probability $\varepsilon$ in all gates of the given circuit, our model also supports to incorporate different $\varepsilon$ values for different gates in the given circuit. Ideally these $\varepsilon$ values has to come from the device variabilities and manufacturing defects. Each gate in a circuit will have an $\varepsilon$ value selected in random from a fixed range, say 0.005 - 0.05.

We have presented the result in {\bf Fig.~\ref{mf_time}} for $max\_flat$. Here we compare the average and maximum output error probability and run time with $\varepsilon$=0.005, $\varepsilon$=0.05 and variable $\varepsilon$ ranging for 0.005 - 0.05. The notable results are as follows,

\begin{itemize}

\item It can be seen that the output error probabilities for variable $\varepsilon$ are closer to those for $\varepsilon$=0.05 than for $\varepsilon$=0.005 implicating that the outputs are affected more by the erroneous gates with $\varepsilon$=0.05. 

\item The run time for all the three cases are almost equal, thereby indicating the efficiency of our model.

\end{itemize}

\section{Conclusion}
\label{conclusion}

We have proposed a probabilistic model that computes the exact maximum output error probabilities for a logic circuit and map this problem as maximum {\it a posteriori} hypothesis of the underlying joint probability distribution function of the network. We have demonstrated our model with standard ISCAS and MCNC benchmarks and provided the maximum output error probability and the corresponding worst-case input vector. We have also studied the circuit-specific error bounds for fault-tolerant computing. The results clearly show that the error bounds are highly dependent on circuit structure and computation of maximum output error is essential to attain a tighter bound. 

Extending our proposed algorithm one can also obtain a set of, say N, input patterns which are highly likely to produce an error in the output. 
Circuit designers will have to pay extra attention in terms of input redundancy for these set of vulnerable inputs responsible for the high end of error spectrum. We are already working on the stochastic heuristic algorithms for both average and maximum error for mid-size benchmarks where exact algorithms are not tractable. This work should serve as a baseline exact estimate to judge the efficacy of the various stochastic heuristic algorithms that will be essential for circuits of higher dimensions. Our future effort is to model the gate error probabilities derived from the physics of the device and fabrication methods; model delay faults due to timing violations; model variability in the error probabilities.

\end{document}